\documentclass[prb, showpacs,amstext,amsxtra,amssymb,latexsym,bbm,twocolumn]{revtex4}
\usepackage{graphicx,epsf}

\usepackage{amsmath,float}
\usepackage{amsfonts}
\usepackage{amsthm}
\usepackage{amssymb}
\usepackage{multirow}
\usepackage{color}
\usepackage{verbatim}

\newcommand{\K}{{\boldsymbol K}}
\renewcommand{\k}{{\boldsymbol k}}
\newcommand{\G}{{\boldsymbol G}}

\newcommand{\be}{\begin{equation}}
\newcommand{\ee}{\end{equation}}
\newcommand{\bs }{\boldsymbol }

\renewcommand{\a}{{\boldsymbol a}}

\newcommand{\D}{{\boldsymbol D}}

\newcommand{\q}{{\boldsymbol q}}

\newcommand{\R}{{\boldsymbol R}}
\renewcommand{\r}{{\boldsymbol r}}

\newcommand{\ep}{\epsilon}

\renewcommand{\d}{{\boldsymbol \delta}}

\begin{document}

%************************************************************************

\title{Artificial graphenes:   Dirac matter beyond condensed matter}

       \author{Gilles Montambaux}
\email{gilles.montambaux@u-psud.fr}
\address{Laboratoire de Physique des Solides, CNRS, Universit\'e Paris-Sud, Universit\'e Paris-Saclay, F-91405 Orsay, France}

\date{\today}

%file : SDH.tex

\begin{abstract}
After the discovery of graphene and its  many fascinating properties, there has been a growing interest for the study of “artificial graphenes”. These are totally different and novel systems which bear exciting similarities with graphene. Among them are lattices of ultracold atoms, microwave or photonic lattices, “molecular graphene” or new compounds like phosphorene. The advantage of these  structures is that they serve as new playgrounds for measuring and testing physical phenomena which may not be reachable in graphene, in particular: the possibility of controlling the existence of Dirac points (or Dirac cones) existing in the electronic spectrum of graphene, of performing interference experiments in reciprocal space, of probing geometrical properties of the wave functions, of manipulating edge states, etc.
These cones, which describe the band structure in the vicinity of the two connected energy bands, are characterized by a topological “charge”. They can be moved in reciprocal space by appropriate modification of external parameters (pressure, twist, sliding, stress, etc.). They can be manipulated, created or suppressed under the condition that the total topological charge be conserved.
In this short review, I discuss several aspects of the scenarios of merging or emergence of Dirac points as well as the experimental investigations of these scenarios in condensed matter and beyond.

\end{abstract}
\maketitle

\section{Introduction}
\label{sect.intro}

\begin{figure}[htbp!]
\begin{center}
\includegraphics[width=8.5cm]{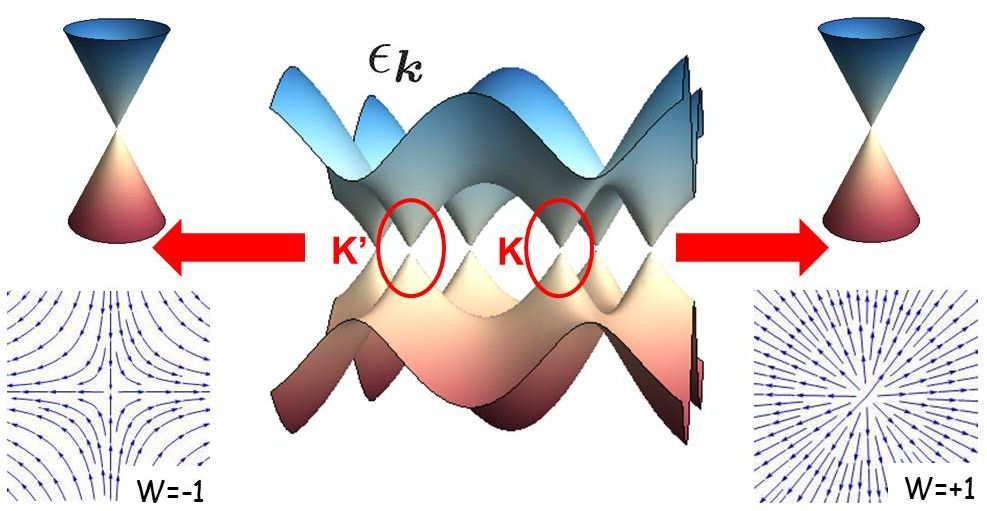}
\end{center}
\caption{\small Top: the energy spectrum $\ep(\k)$ exhibits two inequivalent points K and K' where the dispersion relation is linear. Bottom: representation of the vector field $(\cos \phi_\k, \sin \phi_\k)$ where $\phi_\k$ is the relative phase between the two components of the wave function. The winding of the wave function around K and K' is characterized by a winding number $w=\pm 1$.}
\label{fig:spectre-graphene-dirac}
\end{figure}

  Graphene is a two-dimensional solid constituted by a single layer of graphite, in which the carbon atoms are arranged in a regular honeycomb structure. One of the main interests of this "wonder material"\cite{Rise,grapheneRev} isolated in 2004 is its very unusual electronic spectrum: Fig.\ref{fig:spectre-graphene-dirac} presents the energy $\ep(\k)$ of an electronic state as a function of the electron momentum $\k$, a two-dimensional vector of the reciprocal lattice. As will be reminded in section \ref{sect.reminder}, it consists in two bands touching  at two singular points, named $\K$ and $\K'$, of the Brillouin zone. Around these points the spectrum is linear. These singularities are well described  by the Dirac equation for massless particles in two dimensions. For this reason, these special points of the reciprocal space are usually called {\it Dirac points} and the associated cones are called {\it Dirac cones}. The term of {\it Dirac matter} is frequently used to characterize newly discovered condensed matter systems  having electronic properties described by a Dirac-like equation.\cite{Diracmatter}

 Even more important is the structure of the electron wave functions.  Graphene honeycomb lattice consists in a hexagonal Bravais lattice\cite{remark1} with an elementary pattern made of two inequivalent sites named $A$ and $B$ (fig.\ref{fig:vecteurs}).
  Therefore in the tight-binding approximation, the wave function has  two components corresponding to its amplitude  on $A$ and $B$ sites. The relative phase between these two
components depends on the position of the momentum $\k$ , that is the position of the state in the Brillouin zone, see section \ref{sect.reminder}. Near the Dirac points, this phase has a special structure: it winds around each point and the winding is opposite ($ 2 w \pi$ with $w=\pm 1$) for the two Dirac points (fig.\ref{fig:spectre-graphene-dirac}).
Since the winding is a topological property, the Dirac points behave as topological charges in reciprocal space. These two charges $q$ have opposite signs. It is then natural to ask the question whether these two charges may be moved and even be annihilated. The theoretical and experimental answers to this question are the subject of this work. It has been proposed that the merging of a pair of such Dirac points follows a universal scenario, in the sense that it does not depend on the microscopic problem considered, it describes as well the dynamics of electrons as other waves like microwaves, light or matter waves, that is far beyond the field of graphene or even condensed matter.\cite{Montambaux:09}

Such annihilation of Dirac points is not reachable in graphene because it would imply a large deformation of the lattice impossible to achieve.\cite{Pereira,Goerbig:08} However it may be observed in various different physical systems, including several beyond condensed matter,  which share common properties or interesting analogies with graphene. Such systems are now called "artificial graphenes". The list of such systems is now very long and has been reviewed in a recent paper.\cite{Polini:13} In the present paper, we will not address the huge literature on the subject but only restrict ourselves to several realizations of artificial graphenes in which the motion and merging of Dirac points has been achieved and studied.
\medskip

The outline of this  paper is as follows: we first start with a brief introduction to the electronic properties of graphene.
 Then we show that the motion and merging of Dirac points follows a universal scenario with a peculiar spectrum at the merging transition: it is quadratic along one direction and linear in the perpendicular direction. This peculiar structure of the spectrum, named semi-Dirac, is accompanied by the annihilation of the windings numbers attached to the Dirac points, signature of the topological character of the transition\cite{Volovik}. In the next sections, we consider several experimental works in which this transition has been observed. We emphasize that each experimental situation is not a simple reproduction of graphene physics,  but either allows for the investigation of situations not reachable in graphene or opens new directions of research proper to this situation.
 We conclude by a generalization to various mechanisms of merging of Dirac points.

\section{Brief reminder about graphene}
\label{sect.reminder}

\begin{figure}[htbp!]
\begin{center}
\includegraphics[width=5cm]{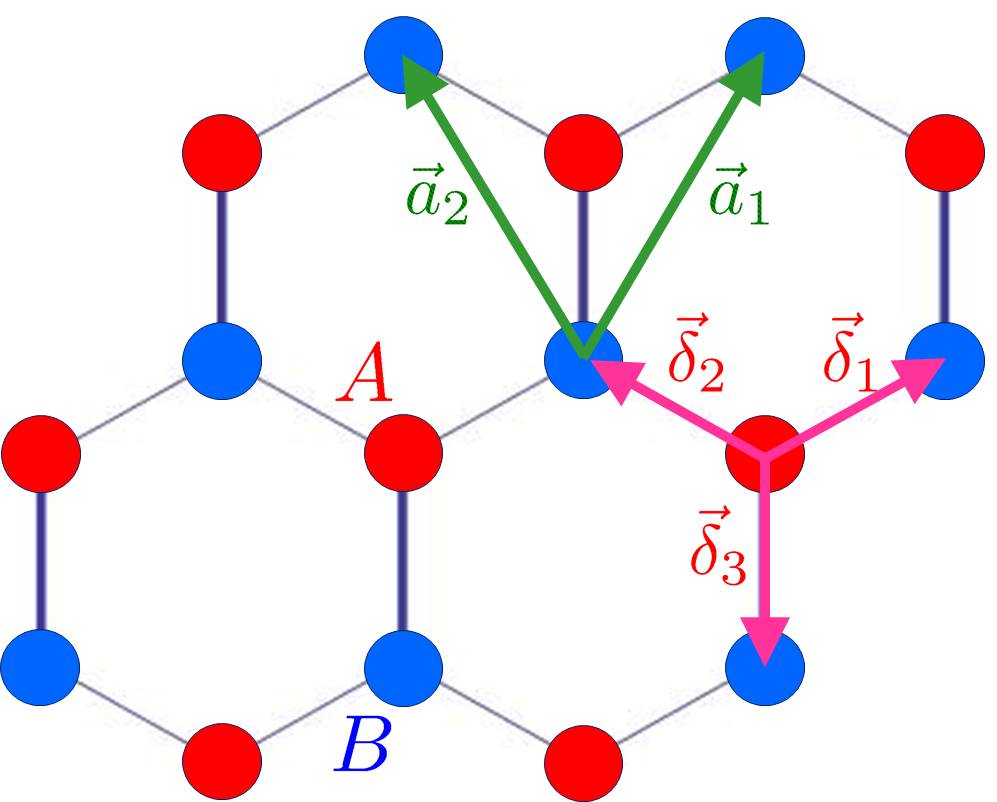}
\end{center}
\caption{\small The "honeycomb" lattice consists in an hexagonal Bravais lattice and two atoms per unit cell. We have represented the elementary vectors $\a_1, \a_2$ of the Bravais lattice and the three vectors ${\bs \delta_1}, {\bs \delta_2}, {\bs \delta_3}$ connecting one site to its nearest neighbours.}
\label{fig:vecteurs}
\end{figure}

\begin{figure}[htbp!]
\begin{center}
\includegraphics[width=9cm]{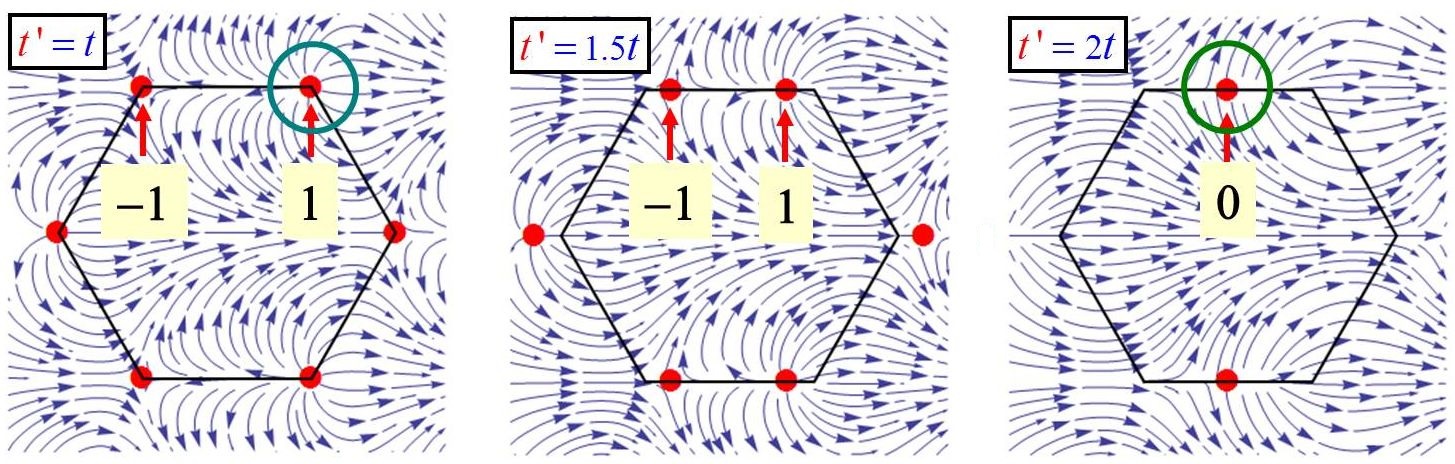}
\end{center}
\caption{\small Evolution of the dependence $\phi_\k$ from the isotropic case $t'=t$ to the merging topological transition $t'=2 t$. At the transition, the opposite winding numbers annihilate. These figures represent the vector field $(\cos \phi_\k, \sin \phi_k)$.}
\label{fig:winding}
\end{figure}

The valence and conduction bands of graphene are described by a simple tight-binding Hamiltonian

\be {\cal H}= - t \sum_{j,j'} | \varphi_j^A \rangle \langle \varphi_{j'}^B | + h. c.  \label{HTB} \ee
where the parameter $t$ characterizes the hopping between nearest neighbouring sites. Additional hoppings between higher nearest neighbours exist but are   omitted here since they do not change much the physics and do not affect the geometric structure of the wave functions. Here we choose to write the eigenfunctions in the form
\be \psi_\k(\r)= e^{i \k \cdot \R} u_\k(\r) \ ,  \ee
where the function $u_\k(\r)$ has the periodicity of the real Bravais lattice {\it and} of the reciprocal lattice (see discussion about this second statement in appendix \ref{appendixA})
 \be
  u_\k(\r+\R) =  u_\k(\r) \qquad ,  \qquad  u_{\k+\G}(\r)= u_\k(\r) \ . \ee
 $\R$ is a vector of the Bravais lattice and  $\G$ is a vector of the reciprocal lattice. In the tight-binding approximation, the cell-periodic part  $u_\k$ is only defined on $A$ and $B$ sites and it is  solution of ${\cal H_\k} u_\k = E_\k u_\k$
with the effective Hamiltonian
\be {\cal H}_\k= \left(
                   \begin{array}{cc}
                     0 & f_\k \\
                     f^*_\k & 0 \\
                   \end{array}
                 \right)
               \label{Heff}  \ee
            and
\be f_\k  = - t( 1 + e^{i \k \cdot \a_1} + e^{i \k \cdot \a_2} ) \ ,  \label{fbaseI}  \ee
where $\a_1$ and $\a_2$ are elementary vectors of the Bravais lattice (fig.\ref{fig:vecteurs}).
The solutions are
\be u_\k={1 \over \sqrt{2}} \left(
                              \begin{array}{c}
                                1 \\
                              \pm e^{i \phi_\k} \\
                              \end{array}
                            \right)
                            \ee
                     with the relative phase $\phi_\k= \arg(f^*_\k)$. They are associated to the energies  $\ep_\k= \pm | f_\k |$ represented in fig.\ref{fig:spectre-graphene-dirac}.
                     The spectrum exhibits  two bands touching linearly in two inequivalent points of the Brillouin zone. In the vicinity of these two "Dirac points",   the Hamiltonian  (\ref{Heff}) takes the linear form:
              \be {\cal H}_\k= c \left(
                   \begin{array}{cc}
                     m c & \hbar (\pm q_x - i q_y) \\
                     \hbar (\pm q_x - i q_y) & -m c  \\
                   \end{array}
                 \right)
               \label{HDirac}  \ee where   $\q= \k - \K^{(')}$, $\hbar c = 3 t a /2$  and $m=0$. This is the form of the Dirac equation for massless particles ($m=0$) in two dimensions.              The energy spectrum is linear
                      \be \ep_\q = \pm \hbar c |\q|   \ee
                       The velocity $c$ is $300$ smaller than the speed of light.
The relative phase $\phi_\k$ is plotted in fig.\ref{fig:winding}. The winding number around a contour $C$ in reciprocal space is defined as\cite{Marzari}
%%%%%%%%%%%%%%%%%%%%%%%%%%%%%%%%%%%%%%%%%%%%%%%%%%%%%%%%
\be w(C)= {1 \over 2 \pi } \oint_C \vec \nabla \phi_\k \cdot d\k  \ . \label{WN} \ee
%%%%%%%%%%%%%%%%%%%%%%%%%%%%%%%%%%%%%%%%%%%%%%%%%%%%%%%%

The two Dirac points are  characterized by opposite winding numbers $w= \pm 1$. One important physical signature of this winding structure  concerns the Landau levels spectrum  $\ep_n$ obtained from the semi-classical Onsager relation\cite{Onsager}
\be N(\ep_n)= (n+ \gamma ) {e B \over h} \ , \label{onsager} \ee
where $eB/h$ is the degeneracy of a Landau level and     $N(\ep)$ is the number of states below energy $\ep$, that is the integral of the density of states (These quantities are written by unit area). The phase factor $\gamma= 1/2 - w(C)/2$ contains the usual Maslov index $1/2$ of the harmonic oscillator and the winding number.\cite{Roth,Fuchs:10} In the vicinity of the Dirac points, the spectrum being linear, the density of states is also linear ($= \ep/\hbar^2 c^2$). Application of the Onsager formula leads to a sequence of Landau levels given by \cite{grapheneRev}
\be
\ep_n(B)= \pm c \sqrt{2 n \hbar e B} \qquad , \qquad n  \in  \mathbb{N}  \ . \label{LLgraphene} \ee
 The observation of this  sequence of levels and in particular the existence of a zero energy level were one of the first signatures  not only of the strictly  2D character of graphene but also of the spinorial character of the wave functions and the Dirac-like structure of the electronic equation of motion.
\cite{kim}

\section{Motion and merging of Dirac points}
\label{sect.merging}

\begin{figure}[h!]
\begin{center}
\includegraphics[width=7cm]{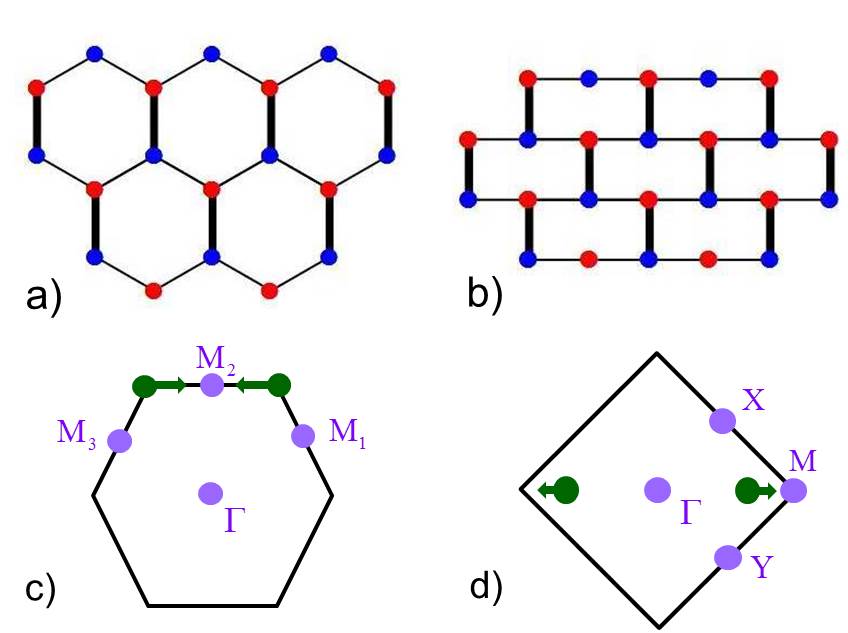}
\end{center}
\caption{\small Top: Honeycomb (a) and brickwall (b) lattices.  Bottom: position (green dots) of the Dirac points for the isotropic honeycomb and brickwall lattices. Their merging can occur at one of the four TRIMs pictured (purple dots) in the first Brillouin zones for the honeycomb (c) and the brickwall (d) lattices.}
\label{fig:graphene-brickwall}
\end{figure}
\subsection{A toy model}
\begin{figure}[htbp!]
\begin{center}
\includegraphics[width=7cm]{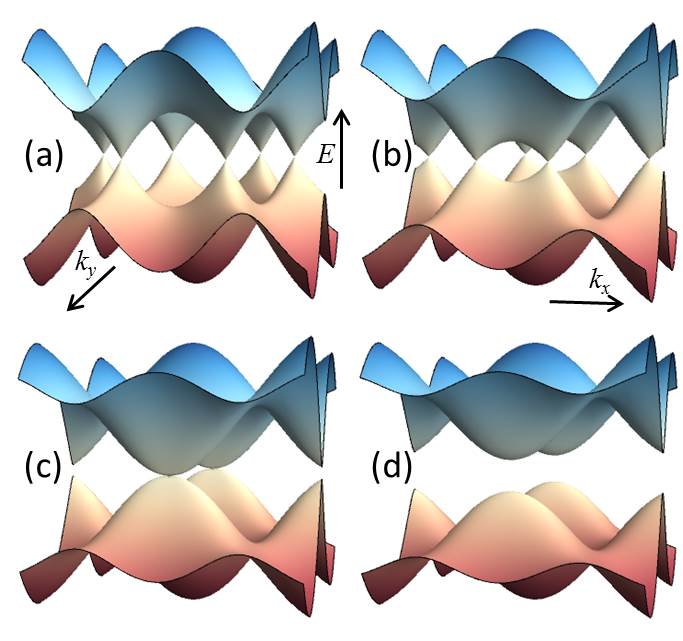}
\end{center}
\caption{\small Evolution of the energy spectrum for the toy model (\ref{Heff},\ref{fdektoy}), with $\beta = t'/t= 1 (a), 1.5 (b), 2 (c), 2.5 (d)$. At  the critical value $\beta=2$, the spectrum is semi-Dirac.}
\label{fig:merging-graphene}
\end{figure}

It is first instructive to consider a toy model which consists in a slight variation of the graphene model.\cite{Montambaux:09,Hasegawa1,Dietl,Guinea:08} One of the three hopping parameters introduced in eq.(\ref{fbaseI}) is now fixed to a different value $t' > t$ (the vertical bars in fig.\ref{fig:graphene-brickwall}). The function $f_\k$ defined above is now given by
\be f_\k= -(t'+ t e^{i \k \cdot \a_1} +t e^{i \k \cdot \a_2}) \ .   \label{fdektoy} \ee
  The position of the Dirac points being given by $f_\k=0$, it is easy to check, and this is shown of fig.\ref{fig:merging-graphene}, that when the parameter $t'$ increases, the Dirac points move until they merge at the M$_2$ point (as defined in fig.\ref{fig:graphene-brickwall}-c). This happens for the critical value of the merging parameter $\beta \equiv t'/t=2$. Above this value, $f_\k=0$ has no solution, implying the existence of a gap in the energy spectrum. The evolution of the spectrum is shown in fig.\ref{fig:merging-graphene}: when $t'$ increases, the two Dirac points tend to approach each other, the velocity along the merging line decreases until it vanishes at the transition $t'=2 t$. At the transition, the spectrum is quite interesting, it is quadratic along the direction of merging, but stays linear along the perpendicular direction. More precisely, it has the form
\be \ep(\q)= \pm \sqrt{ \left({q_x^2 \over 2 m_*}\right)^2 + c^2 q_y^2} \ee
   where $m_*=2/(3 t a^2)$ and $c= 3 t a$ ($\hbar=1$).
   This hybrid spectrum, massive-massless, has been baptized "semi-Dirac".  \cite{Pickett}
   This critical spectrum corresponds to a {\it topological Lifshitz transition}, since the two winding numbers associated to the Dirac points have annihilated and the winding number around the point M is now $0$. At this point and in its vicinity, the thermodynamic and transport properties are quite new.\cite{Montambaux:09,Dietl,Adroguer} The density of states associated to this hybrid spectrum scales  as $\sqrt{\ep}$ and since the winding number is zero, the Onsager-Roth formula leads to a dependence of the Landau levels of the form $\ep_L(B) \propto [(n+1/2) B]^{2/3}$.\cite{Dietl}

\subsection{A universal Hamiltonian}

A natural question now arises whether this toy model is generic of more general situations. We have shown that this scenario of merging is actually quite general.\cite{Montambaux:09} The demonstration is as follows. Consider a general situation where objects (these can be atoms, molecules, classical wave scatterers, semiconducting micropillars, etc.) on a lattice are described by a  Hamiltonian whose general structure is given in eq.(\ref{Heff}).
The function $f$ describes the coupling between $A$ and $B$ sites. Quite generally, the function $f_\k$  can be written in the form (this function has the periodicity of the Bravais lattice, see Appendix \ref{appendixA}):
\be f_\k= \sum_{m,n} t_{mn} e^{i \k \cdot \R_{mn}} \ , \label{fgeneral} \ee  where  $\R_{mn}$  are vectors of a Bravais lattice. For example in the previous toy model, we have simply $t_{00}=t'$, $t_{01}=t_{10}=t$ and other parameters $t_{mn}=0$. In general, depending on the values of these parameters, there may be either no Dirac point,   one or several pairs $(\D,-\D)$ of Dirac points. The merging of two partners necessarily occurs when $\D= - \D \ (\text{mod} \, \G^{pq})$ where $\G^{pq} = p \a^*_1 + q \a^*_2$ is a reciprocal lattice vector. Therefore merging   occurs  at points $\D^{pq}=\G^{pq}/2$. There are four inequivalent (they cannot be deduced from each other by a translation of a reciprocal lattice vector)  such points called Time Reversal Invariant Momenta (TRIM) corresponding to  $(p,q)= (0,0), (0,1), (1,0)$ or $(1,1)$. The merging of a pair of Dirac points occurs at such a point $\D^{pq}$ when
the parameter $f_{\D^{pq}}= \sum_{m,n} t_{mn} e^{i \D^{pq} \cdot \R_{mn}}$ vanishes.  Since the product $ \G^{pq} \cdot \R_{mn}= 2 \pi (pm+qn)$, this parameter  $f_{\D^{pq}}= \sum_{m,n} t_{mn} \beta^{pq}_{mn} $, where
  we have defined  $\beta^{pq}_{mn}=     (-1)^{p m + q n}$.
In the previous example for graphene, merging at the M$_2$ point ($p=1,q=1$) occurs when the parameter $f_{\D^{pq}} =t_{00}-t_{01}- t_{10}=t'-2 t$ vanishes.

\begin{figure}[htbp!]
\begin{center}
\includegraphics[width=8cm]{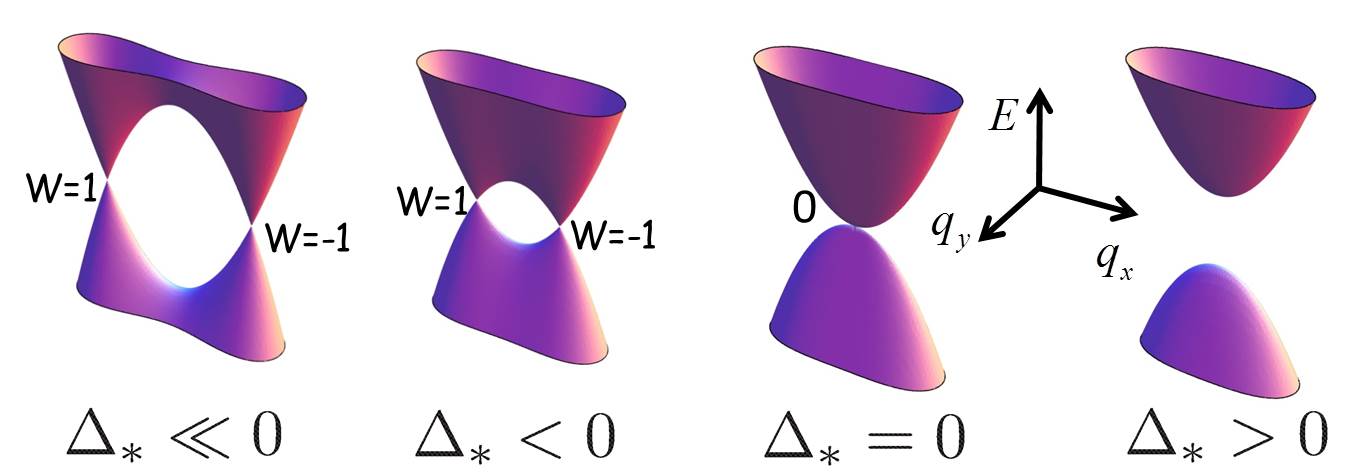}
\end{center}
\caption{\small Universal scenario  for the merging transition between the Dirac phase ($\Delta_* <0$) and the gapped phase ($\Delta_* >0$).}
\label{fig:merging}
\end{figure}

Let us now consider the vicinity of this merging point by an expansion of the wave vector. Writing $\k = \D^{pq} + \q$, we find

\begin{eqnarray} f_{\D^{pq} + \q} &=& f_{\D^{pq}} +  i \q \cdot \sum_{mn} \beta^{pq}_{mn} t_{mn} \R_{mn} \nonumber \\ &-& {1 \over 2}  \sum_{mn} \beta^{pq}_{mn} t_{mn} (\q \cdot \R_{mn})^2 + \cdots \ .  \end{eqnarray}
%%%%%%%%%%%%%%%%%%%%%%%%%%%%%%%%%%%%%%%%%%%%%%%%%%%%%%%%
The linear term is {\it purely imaginary}. It defines a direction $\hat y$ and a velocity
%%%%%%%%%%%%%%%%%%%%%%%%%%%%%%%%%%%%%%%%%%%%%%%%%%%%%%%%
\be c^{pq}= \sum_{m,n} \beta^{pq}_{mn} t_{mn}   R_{y,mn} \ . \ee
%%%%%%%%%%%%%%%%%%%%%%%%%%%%%%%%%%%%%%%%%%%%%%%%%%%%%%%%
Once this local axis is defined, consider the quadratic terms. They are of the form $q_x^2, q_x q_y, q_y^2$ where $\hat x$ is defined has the direction perpendicular to $\hat y$. At lowest order, we keep only the term $q_x^2$.
Then the function $f_{\D^{pq}+ \q}$ has the expansion
%%%%%%%%%%%%%%%%%%%%%%%%%%%%%%%%%%%%%%%%%%%%%%%%%%%%%%%%
\be
f_{\D^{pq}+ \q}=f_{\D^{pq}} + i c^{pq} q_y + {q_x^2 \over 2 m^{pq} }  \ .  \ee
%%%%%%%%%%%%%%%%%%%%%%%%%%%%%%%%%%%%%%%%%%%%%%%%%%%%%%%%
Of course, the parameters $c^{pq}$,  $m^{pq}$  and $\Delta^{pq}$ depend on the position $(p,q)$ of the merging point, and they will be now simply denoted by $c$,  $m_*$ and $\Delta_*$. The effective mass $m_*$ and the merging parameter $\Delta_*$ are given by
%%%%%%%%%%%%%%%%%%%%%%%%%%%%%%%%%%%%%%%%%%%%%%%%%%%%%%%%
\begin{eqnarray} {1 \over m_*}&=& - \sum_{mn}  \beta^{pq}_{mn}  t_{mn} R^2_{x,mn} \ ,  \label{mstar} \\
\Delta_* &=& \sum_{mn}  \beta^{pq}_{mn}  t_{mn}  \label{Deltastar} \ .
\end{eqnarray}
%%%%%%%%%%%%%%%%%%%%%%%%%%%%%%%%%%%%%%%%%%%%%%%%%%%%%%%%
We conclude that, in the vicinity of the merging point at a TRIM, the Hamiltonian as the {\it universal form}:
%%%%%%%%%%%%%%%%%%%%%%%%%%%%%%%%%%%%%%%%%%%%%%%%%%%%%%%%
\be {\cal H}(\q)= \left(
                    \begin{array}{cc}
                      0 & \Delta_* + {q_x^2 \over 2 m_* }+ i c q_y  \\
                     \Delta_* + {q_x^2 \over 2 m_* } - i c q_y  & 0 \\
                    \end{array}
                  \right)
                \label{HU}  \ee
%%%%%%%%%%%%%%%%%%%%%%%%%%%%%%%%%%%%%%%%%%%%%%%%%%%%%%%%
                  where the parameters depend on the position of the merging point.  In addition to being independent of the details of the microscopic parameters, we stress that this is the minimal Hamiltonian which describes the coupling between two Dirac points of opposites charges and their merging.\cite{Montambaux:09}  The dispersion relation  given by
                  \be \ep(\q)= \pm \sqrt{ \left( \Delta_* +{q_x^2 \over 2 m_*}\right)^2 + c^2 q_y^2} \ .  \ee
%%%%%%%%%%%%%%%%%%%%%%%%%%%%%%%%%%%%%%%%%%%%%%%%%%%%%%%%
                  is shown in fig.\ref{fig:merging}.
When $\Delta_*<0$, it describes a pair of   massless  Dirac points with opposite winding numbers.\cite{Montambaux:09} The two Dirac points   at positions $\pm \sqrt{- 2 m^* \Delta_*}$
  are separated by a saddle point (Van Hove singularity) at energy $|\Delta_*|$. At the critical value $\Delta_*=0$, the spectrum has the hybrid "semi-Dirac" form, quadratic in the $\hat x$ direction and linear in the $\hat y$ direction. The density of states associated to this hybrid spectrum scales as $\sqrt{\ep}$. When $\Delta_*>0$, a gap opens.
The Hamiltonian (\ref{HU}) describes a continuous cross-over between the limit of uncoupled valleys corresponding to the situation in graphene and  the merging of the Dirac points. It allows for a continuous description of the coupling between valleys which increases as the energy of the saddle decreases.
\medskip

\begin{figure}[h!]
\begin{center}
\includegraphics[width=6cm]{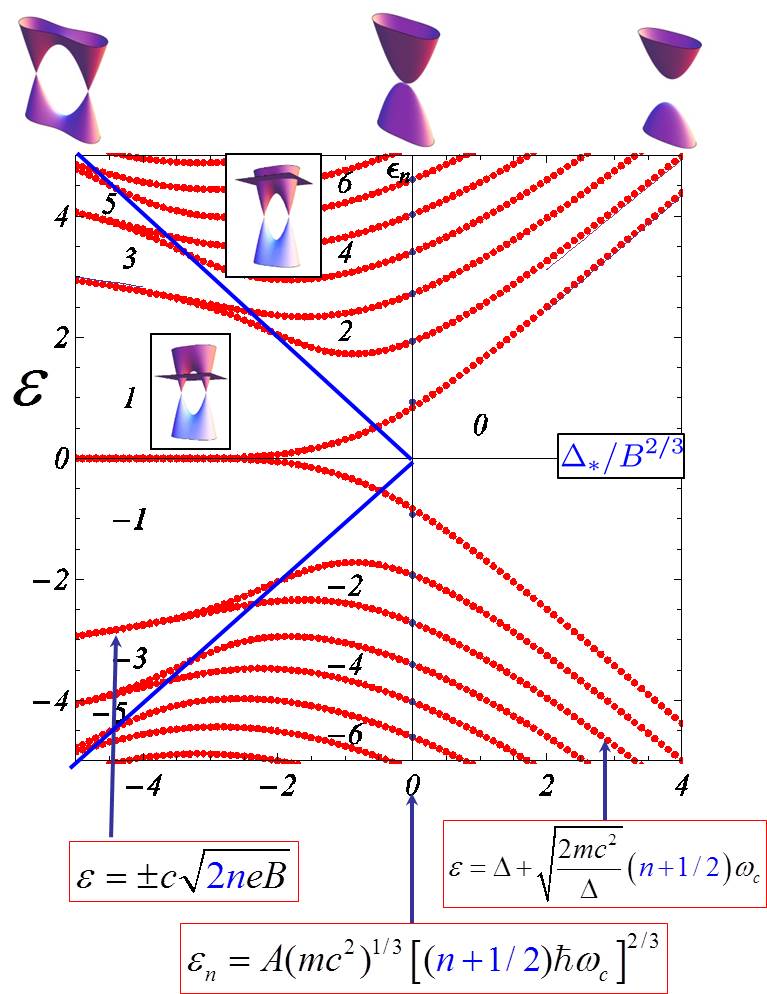}
\end{center}
\caption{\small Evolution of the Landau spectrum as a function of the parameter $\Delta^*/B^{2/3}$. The sequence of Hall numbers is indicated in the gaps. The blue line corresponds to the energy of the saddle point separating Dirac cones. For more details, see ref.\onlinecite{Montambaux:09}.}
\label{fig:landau-spectrum}
\end{figure}

In refs.\onlinecite{Dietl,Montambaux:09,Adroguer}, we have studied the thermodynamic and transport properties of this  Hamiltonian, in particular the spectrum in a magnetic field. The evolution of the Landau levels from the Dirac phase $\Delta_* <0$ to the gapped phase $\Delta_*>0$ is shown in fig.\ref{fig:landau-spectrum}. In the Dirac phase, the Landau levels vary as $\sqrt{n B}$ (eq.\ref{LLgraphene}) with a double valley degeneracy. When the energy of the Landau levels reaches the energy of the saddle point $|\Delta_*|$, the two-fold valley degeneracy is lifted. At the critical value $\Delta_*= 0$, the energy levels scale as $[n+1/2)B]^{2/3}$, a consequence of the $\sqrt{\ep}$ dependence of the density of states and of the cancellation of the winding number. Above the transition, in the gapped phase, one recovers usual Landau levels varying linearly with the field.
\medskip

\begin{figure}[h!]
\begin{center}
\includegraphics[width=6cm]{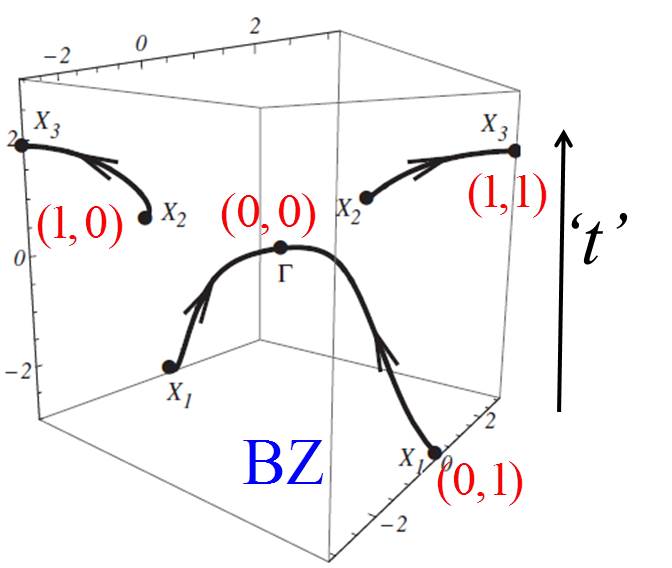}
\end{center}
\caption{\small Motion of Dirac points when varying the parameters of eq.(\ref{tij}). Here $'t'$ represents a variation of hopping parameters which is not detailed here. See ref.\onlinecite{Montambaux:09} for the choice of parameters.}
\label{fig:mergings}
\end{figure}

In ref.\onlinecite{Montambaux:09} it  was proposed  a slightly more general toy model than the one presented above. It consists of a    lattice  with four different hopping parameters  $t_{00}, t_{01}, t_{10}, t_{11}$. In this case, the merging parameter $\Delta_*$ depends on the merging point (one of the four TRIMs) and it is given by eq.(\ref{Deltastar}), that is:
\be \Delta_*=t_{00}+ (-1)^p t_{01} + (-1)^q t_{10} + (-1)^{p+q} t_{11} \label{tij} \ee
%%%%%%%%%%%%%%%%%%%%%%%%%%%%%%%%%%%%%%%%%%%%%%%%%%%%%%%%%%%%%
Put it differently, the merging or emergence of Dirac points is possible at the TRIM $(p,q)$ where this parameter vanishes. Fig.\ref{fig:mergings} shows the motion of the Dirac points between the four different TRIMs under appropriate variation of hopping parameters. We shall see in the next section that this tight-binding model has been achieved experimentally.
The  particular choice of parameters $t_{00}=t'$, $t_{01}=t_{10}=t$, $t_{11}=0$ corresponds to the "brickwall" lattice, very similar to graphene, but with a rectangular symmetry, see Figs. \ref{fig:graphene-brickwall} and \ref{fig:briques}.

The  scenario described by the universal Hamiltonian was proposed in 2009. It was immediately realized that the merging of Dirac points could not be reached in graphene, because it would need an applied strain along one direction of about $23\%$.\cite{Pereira,Goerbig:08}  This scenario  remained considered as a theoretical exercise for a few years. It was predicted to be accessible in the organic conductor $\alpha$-(BEDT-TTF)$_2$I$_3$ under high pressure but has not been observed yet.\cite{Kobayashi} It took a few years to highlight this scenario in various systems emulating the physics of electronic propagation in graphene. These systems are now called "artificial graphenes".
Their interest lies in the strong experimental challenge for their realization and the variety of physical questions they raise. In the next sections, we detail several experiments on artificial graphenes in which the motion and merging of Dirac points have been highlighted.

\section{Cold atoms}
\label{sect.coldatoms}
 \fbox{\parbox{8cm}{\vspace{-3mm}
 \begin{eqnarray} \text{electrons}   & \longrightarrow & \text{atoms} \nonumber \\
\text{atomic lattice}   & \longrightarrow &  \text{optical lattice} \nonumber
\end{eqnarray}
}}
\medskip

\begin{figure}[htbp!]
\begin{center}
\includegraphics[width=3.5cm]{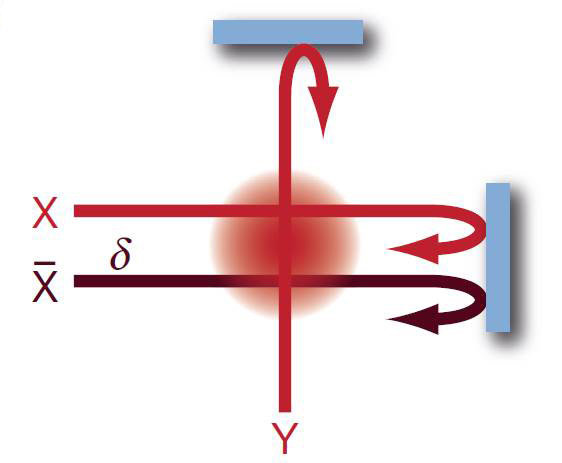}
\includegraphics[width=9cm]{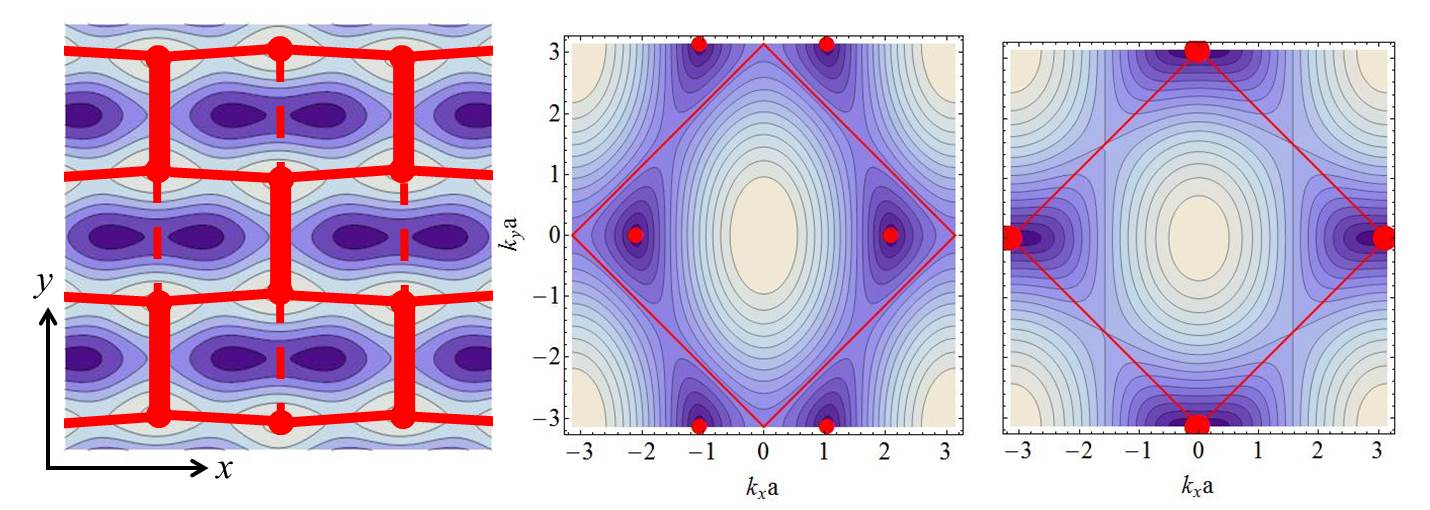}
\end{center}
\caption{\small {Top: Three lasers beams create a two-dimensional potential (see ref.\onlinecite{Tarruell:12} for details). Bottom left: Spatial dependence of this optical potential}  described by a tight-binding model.  The atomic sites corresponds to the minima of the potential (light blue) Tunneling is avoided along the maxima of the potential (dark blue).  The lines represent the hopping amplitudes ($t$ : horizontal lines, $t'$ thick vertical lines. $t''$ dashed vertical lines). Bottom right: isoenergy lines and position of the Dirac points for $t'=t$ and $t'=2 t$ ($t''=0$).}
\label{fig:briques}
\end{figure}

\begin{figure}[htbp!]
\begin{center}
\includegraphics[width=4.5cm]{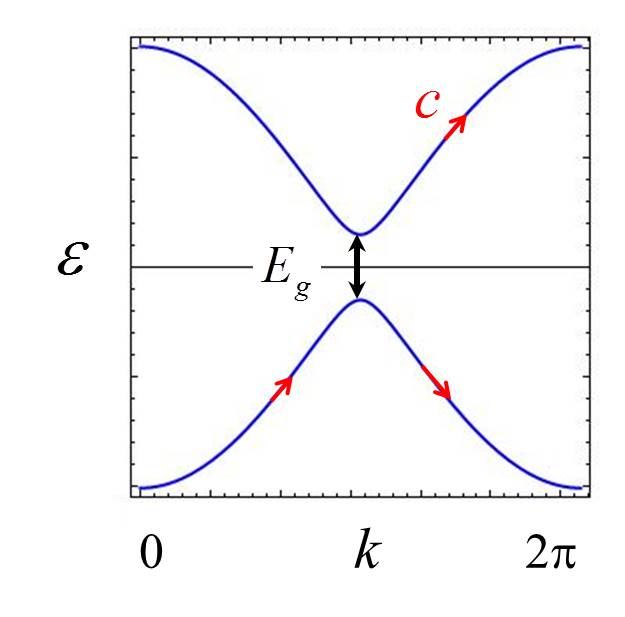}
\end{center}
\caption{\small    A atom initially in the lower band   submitted to a constant force $F$ may tunnel from the lower to the upper band.}
\label{fig:LZ}
\end{figure}

In condensed matter the structure of the atomic orbitals makes it impossible to realize a brickwall lattice. However, such a lattice can be achieved with ultracold atoms trapped in an optical lattice.\cite{Tarruell:12} An appropriate potential profile, schematized in fig.\ref{fig:briques}, is generated with three stationary laser beams.   With appropriate laser intensities  (denoted $V_X$ and $V_{\overline X}$, see fig.\ref{fig:briques} and ref.\onlinecite{Tarruell:12}), the energy spectrum of  the trapped atoms exhibits two bands with a pair of Dirac points, well described by the brickwall tight-binding model with essentially two parameters $t$ and $t'$, which can be tuned by modification of the laser parameters. A third parameter $t''$ may be introduced for a better quantitative modeling. The motion and merging of these points are properly tuned by variation of the laser parameters. In order to probe the spectrum and its evolution, a low-energy cloud of fermionic atoms $^{40}$K  occupies the bottom of the lower band and is submitted to a constant force $F$   inducing a linear time dependence $\hbar k = F t$ of the quasi-momentum $\k$, so that the energy  $\ep(\k)$ oscillates with time. This phenomenon is called a Bloch oscillation.\cite{Bloch,Dahan} If the motion is slow (small $F$), a state $k(t)$ moves adiabatically along the energy band $\ep(k(t))$. If the motion is fast (large $F$) and when two bands become close to each other, an atom initially in the lower band has a finite probability to tunnel to the upper band. Near a Dirac point, the tunneling probability is given by the Landau-Zener-St\"uckelberg  theory and is related to the gap $E_g$ between the two bands:\cite{Zener}
%%%%%%%%%%%%%%%%%%%%%%%%%%%%%%%%%%%%%%%%%%%%%%%%%%%%%%%%%%%%%%%%%%%
\be P_Z= e^{\displaystyle - \pi {(E_g/2)^2 \over c F}}  \ . \label{eq.zener} \ee
%%%%%%%%%%%%%%%%%%%%%%%%%%%%%%%%%%%%%%%%%%%%%%%%%%%%%%%%%%%%%%%%%%%
$F$ is the force which drives the linear time dependence of the wave vector $k$ and $c$ is the velocity (fig. \ref{fig:LZ}).
 By measuring the proportion of atoms transferred to the upper band after a single Bloch oscillation, it is possible to reconstruct the energy spectrum in the vicinity of the Dirac points, therefore to highlight the motion and merging of Dirac points under distortion of the lattice induced by appropriate variations of the laser parameters.
Two geometries are of interest, when the force is applied along or perpendicularly to the merging line (fig.\ref{fig:cold-zener}-a,d).

\subsection{Single Zener tunneling}

\begin{figure}[htbp!]
\begin{center}
\includegraphics[width=8.5cm]{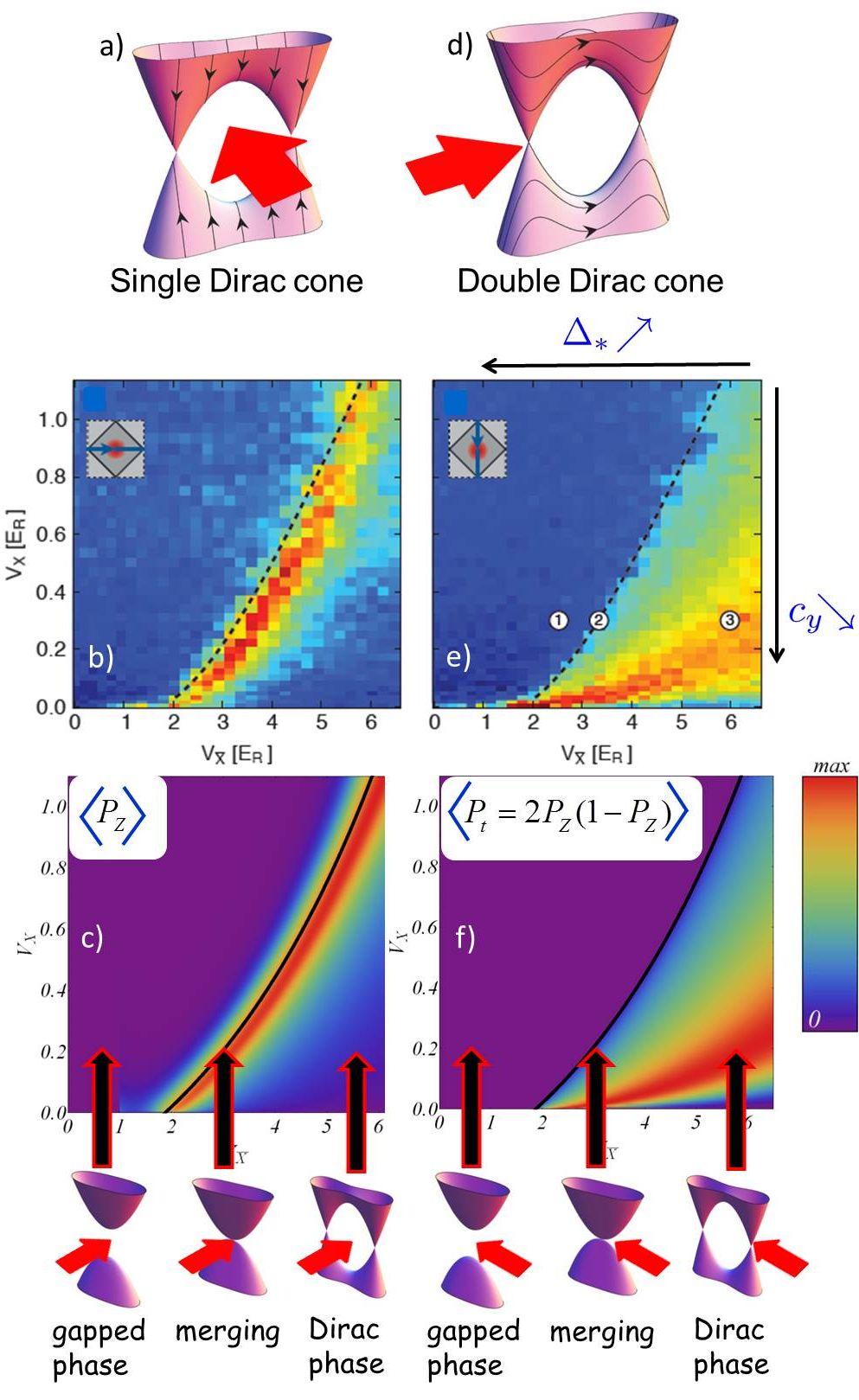}
\end{center}
\caption{\small A low energy fermionic cloud of cold atoms is accelerated towards a pair of Dirac points (a,b). The  probability to tunnel to the upper band depends on the vicinity of the Dirac cones. This probability is represented as a function of the two laser parameters $V_X$ and $V_{\overline X}$, see refs.\onlinecite{Tarruell:12,Lim:12}.   Top figures b,e : experimental results.\cite{Tarruell:12}  The dashed line indicates the merging transition (2) separating the gapped phase (1) and the Dirac phase (3).   Figures c,f : theory.\cite{Lim:12} After a single Zener tunneling event, the probability is maximal close to the merging line (left figures). When a double Zener tunneling, the probability in maximal inside the Dirac phase.}
\label{fig:cold-zener}
\end{figure}

In this first situation, the force $F$ in applied along the $\hat y$-direction perpendicular to the merging line so that the fermionic cloud hits the two Dirac points in parallel (fig.\ref{fig:cold-zener}-a).
An atom in a state with finite $q_x$    performs a Bloch oscillation along a direction of constant $q_x$ and may tunnel into the upper band with the probability

\be P_Z^y= \displaystyle e^{\displaystyle -\pi {(\Delta_*+{q_x^2 \over 2 m^*})^2 \over c_y F}} \ .
\label{PyZ} \ee

In the gapped (G) phase ($\Delta_* >0$), the tunneling probability is vanishingly small. Deep in the Dirac (D) phase ($\Delta_* <0$), when the distance between the two Dirac points $2q_D= 2\sqrt{2 m_* |\Delta_*|}$ is larger than the size of the cloud, the tunnel probability is also small  (fig.\ref{fig:cold-zener}-b). It is maximal close to the merging transition.
In order to simulate the finite width of the fermionic cloud, the probability (\ref{PyZ}) has to be averaged over a finite range of $q_x$. Then the three parameters of the universal Hamiltonian can be related to the tight-binding parameters and then,  via a simple {\it ab initio} calculation,  to the two parameters $(V_X,V_{\overline X})$ of the optical potential to finally  obtain the probability as a function of these two parameters.  As shown in fig.\ref{fig:cold-zener}, we have found that the laser amplitude $V_{\overline X}$ mainly tunes the merging parameter $\Delta_*$ while the laser amplitude $V_X$ essentially tunes the velocity $c_y$. One obtains without adjustable parameter an excellent agreement with the experimental result (compare Figs. \ref{fig:cold-zener}-b-c).\cite{Lim:12} In particular we find that the probability is maximal for $\Delta \simeq -\langle q_x^2 \rangle /2 m_*$ that is {\it inside} the Dirac phase as found experimentally.

\subsection{Double Zener tunneling}

The force $F$ is now applied along the merging direction $\hat x$ so that the fermionic cloud hits the two Dirac points in series (fig.\ref{fig:cold-zener}-d).
Along one Bloch oscillation, an atom of  fixed $q_y$  performs now two Zener transitions in a row, each of them being characterized by the tunnel probability

\be P_Z^x= \displaystyle e^{\displaystyle -\pi  {c_y^2 q_y^2 \over  c_x F}}= \displaystyle e^{\displaystyle -\pi  {c_y^2 q_y^2 \over  F \sqrt{2 |\Delta_*|^2/m_*}}} \ .
\label{PxZ} \ee
Assuming that the two tunneling events are incoherent, the total interband transition probability is

%%%%%%%%%%%%%%%%%%%%%%%%%%%%%%%%%%%%%%%%%%%%%%%%%%%%%%%%%%%
\be P_t^x = 2 P_Z^x (1 - P_Z^x) \ .  \label{Ptx} \ee
%%%%%%%%%%%%%%%%%%%%%%%%%%%%%%%%%%%%%%%%%%%%%%%%%%%%%%%%%%%

In the gapped phase, the tunneling probability is again vanishingly small.
In the Dirac phase, the interband transition probability [Eq. (\ref{Ptx})] is a non-monotonic function of the Zener probability $P_Z^x$, and it is  maximal when $P_Z^x=1/2$.  This explains why the maximum of the tunnel probability  is located well {\it inside} the D phase (red region in Figs.  \ref{fig:cold-zener}-e,f). Again, after averaging over $q_y$, the size of the initial cloud, and relating  the parameters $\Delta_*, m_*$ and $c_y$  to the microscopic parameters $(V_X,V_{\overline X})$, one finds an excellent agreement with the experimental result (compare Figs. \ref{fig:cold-zener}-e-f).

 \subsection{Interferometry in reciprocal space}

 The double Zener configuration is particularly interesting because of possible interference between the  two Zener events. Assuming the phase coherence is preserved, instead of
the probability given by Eq. (\ref{Ptx}), one expects a resulting probability of the form

%%%%%%%%%%%%%%%%%%%%%%%%%%%%%%%%%%%%%%%%%%%%%%%%%%%%%%%%%%%
\be P_t^x = 4 P_Z^x (1 - P_Z^x)  \cos^2 (\varphi/2 + \varphi_d) \ ,  \label{Ptxcoh} \ee
%%%%%%%%%%%%%%%%%%%%%%%%%%%%%%%%%%%%%%%%%%%%%%%%%%%%%%%%%%%
where $\varphi_d$ is  a phase delay, named Stokes phase, attached to the each tunneling event, and $\varphi=\varphi_{dyn}+\varphi_{g}$ is a phase which has two contributions,
a dynamical phase $\varphi_{dyn}$ acquired between the two tunneling events  and basically related to the energy difference between the two energy paths, and a geometric phase $\varphi_{g}$. Whereas the dynamical phase carries information about the spectrum, the geometric phase carries information about the structure of the wave functions.\cite{Shevchenko,Lim:13} If the interference pattern could not been observed in the experiment of ref.\onlinecite{Tarruell:12}, probably because the optical lattice was actually three-dimensional, it is an experimental challenge to access it directly   and to probe the different contributions to the dephasing. More recent experiments with a honeycomb lattice of ultracold atoms have permitted to completely reconstruct the full structure of the wave functions,\cite{Schneider,Tarnowski} as expected theoretically. \cite{Lim:15}

\section{Microwaves}
\label{sect.microwaves}

 \fbox{\parbox{8cm}{\vspace{-3mm}
 \begin{eqnarray} \text{electrons}   & \longrightarrow & \text{microwaves} \nonumber \\
\text{atomic lattice}   & \longrightarrow &  \text{lattice of dielectric resonators} \nonumber
\end{eqnarray}
}}
\medskip

\begin{figure}[htbp!]
\begin{center}
\includegraphics[width=8cm]{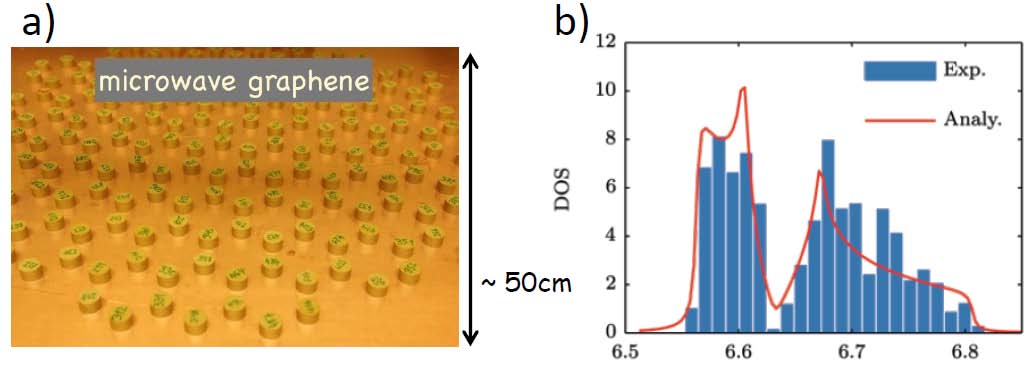}
\includegraphics[width=8cm]{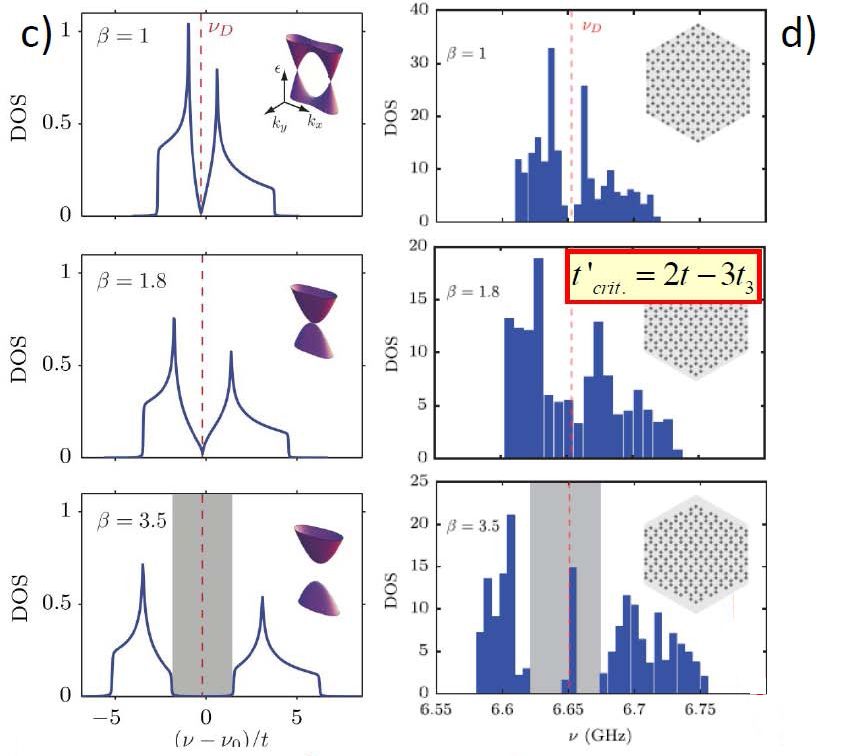}\\
\end{center}
\caption{\small a) Honeycomb lattice of 288 dielectric cylinders. b) Measured density of states well fitted by a tight-binding model with second and third nearest neighbor couplings. c) Expected evolution of the density of states when increasing the anisotropy parameter $\beta=t'/t$. d) Experimental evolution of the density of states. The merging transition occurs when $\beta=2 - 3 t_3/t$. Under strain, new edge states appear at the band center. }
\label{fig:microwave}
\end{figure}

Another artificial graphene has been simulated with a macroscopic setup where cm-size cylindrical  dielectric dots replace carbon atoms and microwaves replace the electrons.\cite{{Bellec:13a}} A few hundred dots are sandwiched between two metallic plates and are arranged in a 2D honeycomb lattice. A microwave is send  through this artificial crystal and is collected by an antenna. The measured signal is related to the local density of states at the place of the antenna. The global density of states is obtained from averaging on the position of the antenna. Therefore this new setup realizes a microwave analog simulator of the quantum propagation of electrons in graphene. The distance between the dots fixes the intensity of the electromagnetic coupling.  The frequency of the microwave (a few GHz) is such that its propagation   is resonant in the dots and is evanescent outside. Its propagation in the lattice is thus well-described by the same tight-binding model as for graphene. Like in graphene, the second and third neighbour couplings are not negligible (here $t_2/t \sim -0.09$, $t_3 \simeq 0.07$). They do not alter the properties of the Dirac cones, but they change the overall shape of the density of states (fig.\ref{fig:microwave}).\cite{Bellec:13b}
 \medskip

 This setup allows for the observation of   effects hardly visible or difficult to explore in graphene like the topological transition. Here it can be reached by fabricating an appropriately uniaxial strained lattice. By doing so, the hopping parameters are modified, essentially the nearest neighbour hopping $t'$ along the strain direction.  The merging parameter $\Delta_*$ is given by  eq.(\ref{Deltastar}) with the additional parameters $t_{11}=t_{1,-1}=t_{-1,1}=t_3$, that is ($p=q=1$, $m=n=1$):
\be \Delta_* = t' - 2 t + 3 t_3 \ , \ee
so that the merging transition is expected for the parameter $t'= 2 t - 3 t_3 \simeq 1.8 t$, as observed experimentally (fig.\ref{fig:microwave}).
\medskip

\begin{figure}[htbp!]
\begin{center}
\includegraphics[width=3cm]{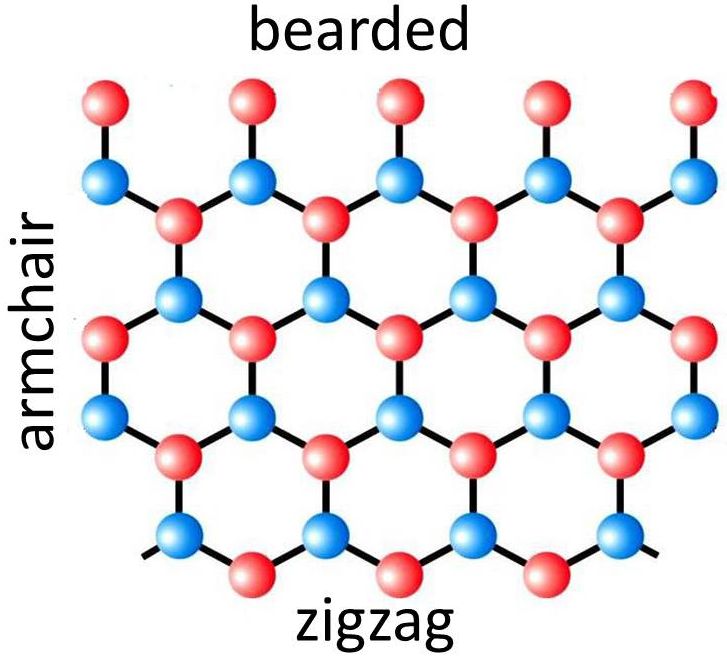}
\includegraphics[width=8.5cm]{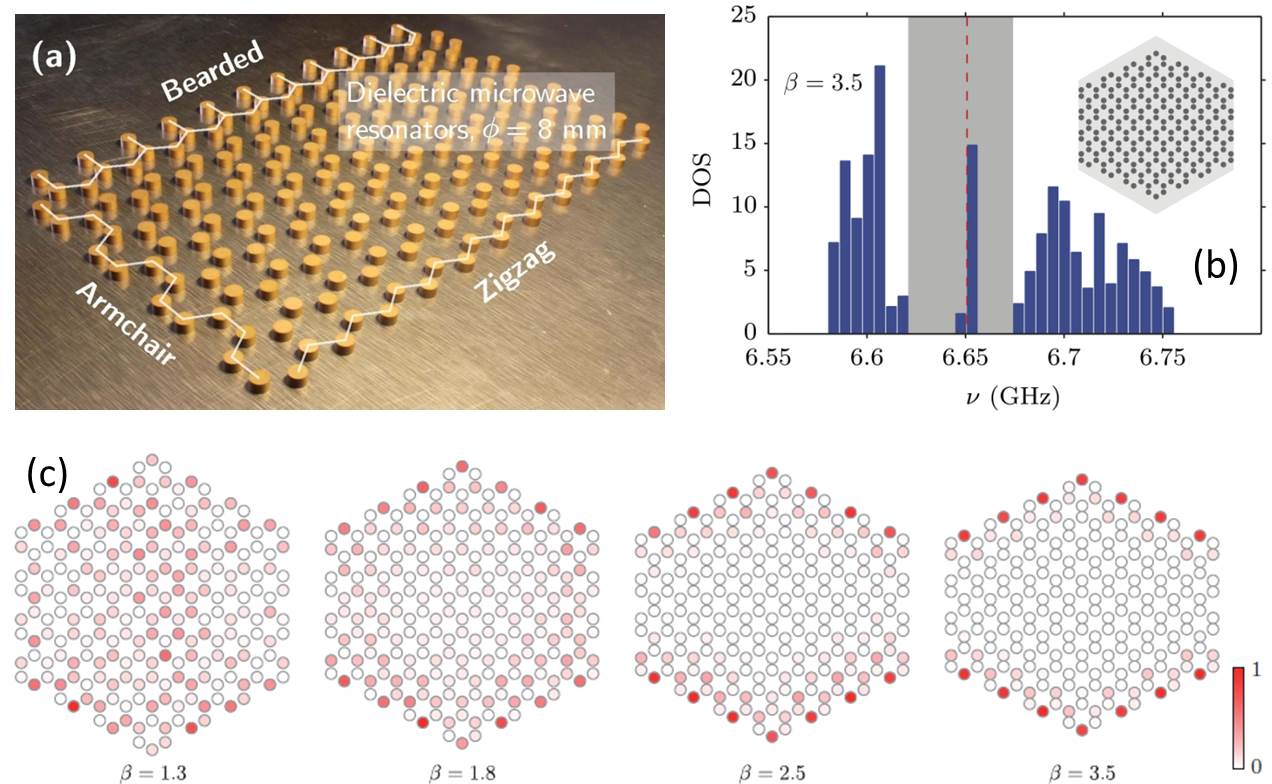}
\end{center}
\caption{\small a) The three types of edges, armchair, zigzag and bearded, are achieved on this sample. b) In this lattice under strain, zero energy localized states appear along the armchair edges. c) Evolution of the detected intensity along the armchair edges, showing the appearance of edge states.\cite{Bellec:13a}}
\label{fig:ac-edges}
\end{figure}

  Moreover, the simplicity of this setup   allows to probe easily   the existence of new exotic states at the edges of the system. The configuration of the edges, well-known to play an important role in graphene but difficult to control, is quite easy to modify here.
  Fig.\ref{fig:ac-edges} shows a lattice of dielectric dots with the three types of edges, armchair, zigzag and bearded. The last type of edge, hard to achieve in condensed matter, is quite easy to build here.
 In this experiment, it has been possible to control the existence of localized states along these different edges, and to control the different types of edge states under applied strain.\cite{Bellec:14} Moreover it is possible to measure, for each state, the spatial repartition of the microwave field along the edges. For example, the experiment of fig.\ref{fig:microwave} was performed on a sample with armchair edges which are known not to support localized edge states. However under strain along an  appropriate direction, it was predicted that zero energy localized states would appear along the armchair edges.\cite{Delplace:11}
  This evolution is seen on Figs.\ref{fig:microwave}-d and \ref{fig:ac-edges}-c.

  \begin{center}
  \begin{figure}[htbp!]
\begin{center}
\includegraphics[width=8.5cm]{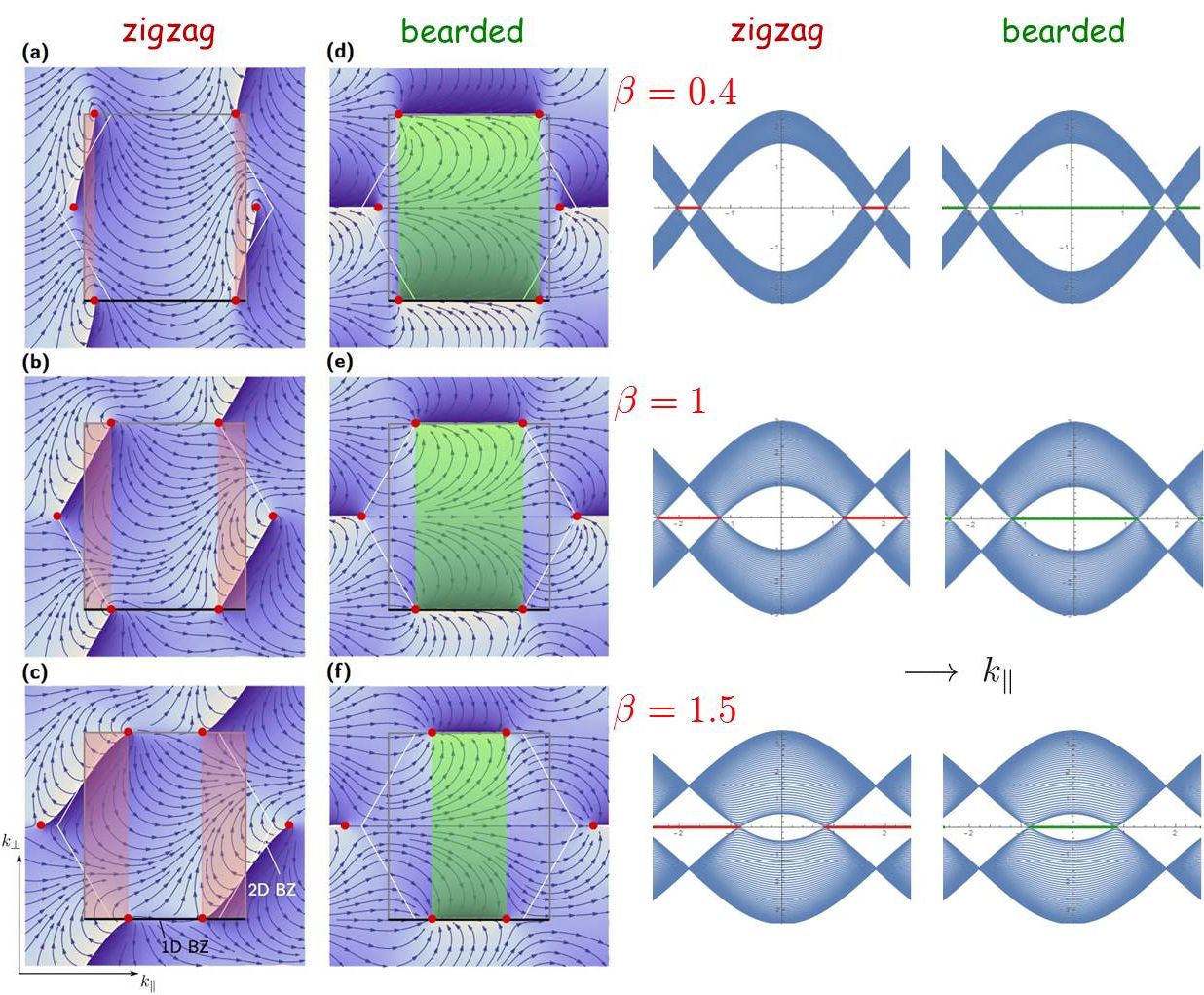}
\end{center}
\begin{center}
\includegraphics[width=8.5cm]{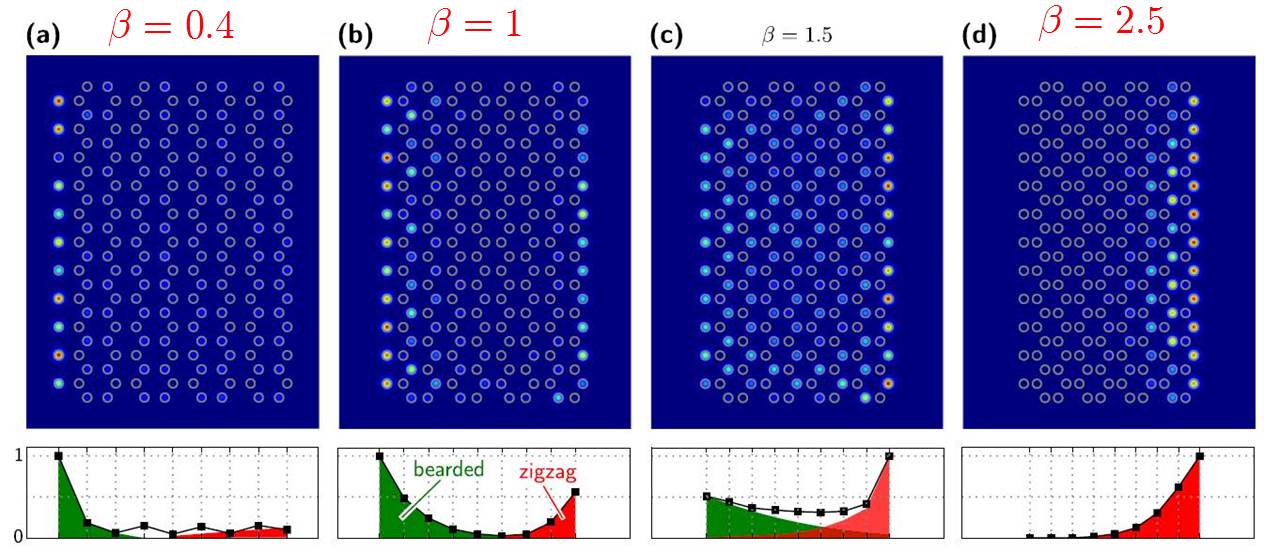}
\end{center}
\caption{\small Top left: winding of the phase $\phi_\k$ in the Brillouin zone.   $w(k_\parallel)$, here $0$ or $1$,  measures the number of edges states. There range of existence is colored in pink (zigzag edge states) and green (bearded edge states). Top right : energy spectrum as a function of $k_\parallel$, the wave vector along the direction of the edge. Zero energy edge states are represented by red and green lines. When $\beta$ increases, new zigzag states appear at the expense of bearded edge states.  This is clearly visible on the experiment (ref. \onlinecite{Bellec:14}) which directly measures the intensity of the microwave at the position of each dot. }
\label{fig:zz-bd}
\end{figure}
\end{center}

  It has been shown that the existence of zero energy edge states can be related to a geometric property of the bulk wave functions encoded in a topological quantity bearing similarity with a Zak phase, a one-dimensional equivalent of the Berry phase.\cite{Delplace:11} First the Hamiltonian, that is the function $f_\k$, has to be written consistently with the type of edge, as explained in ref.\onlinecite{Delplace:11}. Then  the topological quantity
\be w(k_\parallel) = {1 \over 2 \pi} \int {\partial  \phi_\k \over \partial k_\perp}  d k_\perp \ee
measures the winding of the wave function along the direction $\perp$ perpendicular to the edges.  It directly gives the number of edge states with  wave vector $k_\parallel$ along a given edge. Fig.\ref{fig:zz-bd} shows the winding of the phase and values of $k_\parallel$ for which zero energy edge states exists. It is seen that zigzag and bearded states are complementary and that a strain perpendicular to the edges modify the relative proportion of bearded and zigzag states.  When $\beta <1$, new bearded edge states appear at the expense of zigzag edge states. On the opposite, when $\beta >1$, new zigzag edge states appear at the expense of bearded edge states, see bottom fig.\ref{fig:zz-bd}.  A similar experiment has been realized in another "photonic" graphene realized with an array of waveguides arranged in the honeycomb geometry.\cite{Segev:13}

\section{Polaritons}
\label{sect.polaritons}

 \fbox{\parbox{8cm}{\vspace{-3mm}
 \begin{eqnarray} \text{electrons}   & \longrightarrow & \text{polaritons} \nonumber \\
\text{atomic lattice}   & \longrightarrow &  \text{SC micropillars} \nonumber
\end{eqnarray}
}}
\medskip

\begin{figure}[htbp!]
\begin{center}
\includegraphics[width=6.5cm]{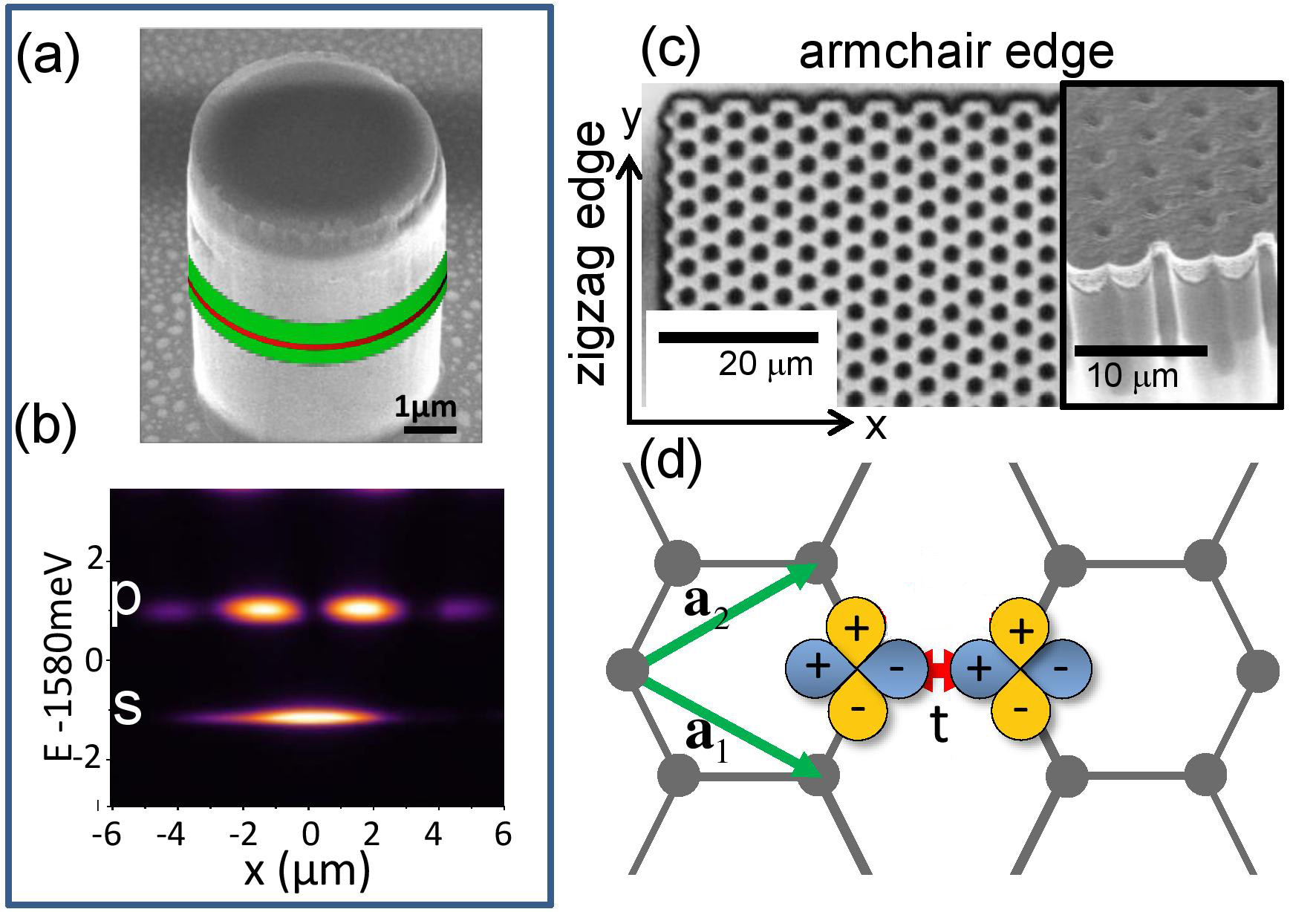}\includegraphics[width=2.5cm]{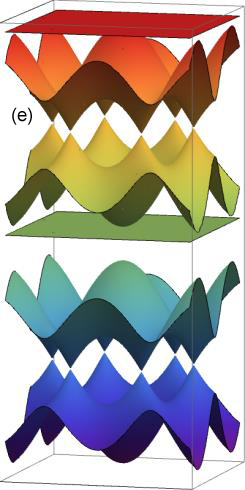}
\end{center}
\caption{\small (a) Polaritons in a micropillar: Light is confined in a GaAs cavity (green) where it interacts with quantum well excitons (red). (b) Photonic states are discrete, like the electronic levels in an atom. (c) Honeycomb lattice of overlapping pillars. Polaritons can propagate in this lattice, like electrons do in graphene. (d) Orbital graphene is obtained by coupling the $p_x$ and $p_y$ states of adjacent sites in the lattice. (e) Band structure exhibiting the $p$-spectrum on top of the $s$-spectrum. The $p$-spectrum resembles the $s$-spectrum, with two additional flat bands (green and red).}
\label{fig:polaritons1}
\end{figure}

Another very interesting setup is a honeycomb lattice of semiconducting (SC) micropillars in which polaritons propagate.\cite{c2n0} These mixed light-matter quasiparticles arise from the strong coupling between electronic excitations (so-called excitons) and photons confined in a semiconductor cavity grown by molecular beam epitaxy (fig.\ref{fig:polaritons1}-a). One such cavity, shaped in the form of a micropillar with lateral dimensions of the order of a few microns, behaves like an artificial atom: its discret energy states (orbitals) are the allowed modes of confined photons (fig.\ref{fig:polaritons1}-b). By exciting the cavity with a laser beam, generated electron-hole pairs  relax to form polaritons that populate the lowest energy levels. The measurement of energy and angle of the emitted photons allow to reconstruct the polariton spectrum.

\begin{figure}[h!]
\begin{center}
\includegraphics[width=6cm]{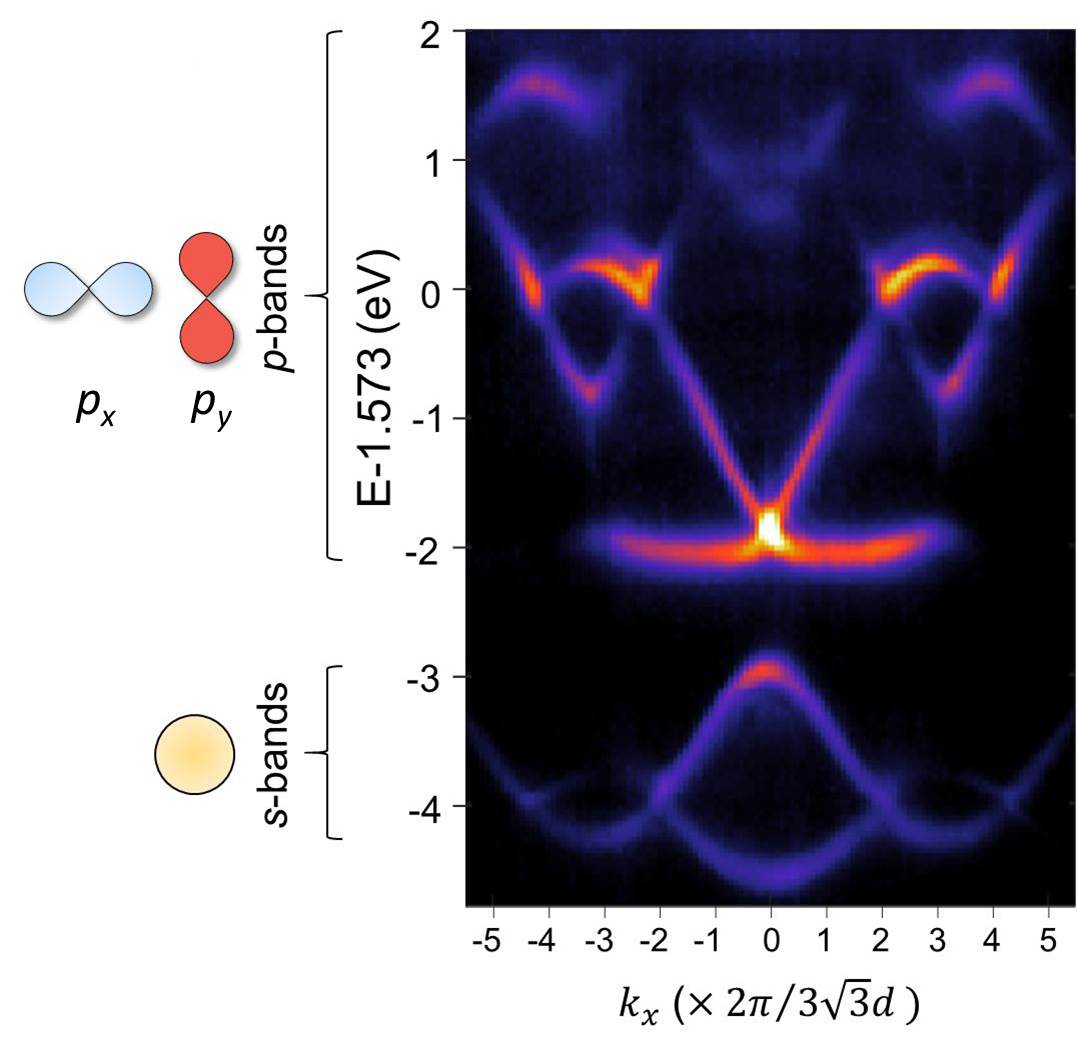}
\end{center}
\caption{\small This photoluminescence spectrum shows the two lowest bands of $s$ symmetry and the higher $p$-bands. The lowest flat band is visible while the upper flat band is hardly visible because it is hybridized with higher orbitals.}
\label{fig:s-p-spectra}
\end{figure}

By coupling these cavities to make a honeycomb lattice (fig.\ref{fig:polaritons1}-c), one obtains an artificial structure analogue to graphene: the micropillars play the role of carbon atoms, and polaritons which propagate between the pillars play the role of electrons in graphene. The measurement of the dispersion relation of the polaritons by photoluminescence is quite analogous to ARPES experiments in condensed matter. One of the advantages of this system is the possibility to probe the higher orbitals of the pillars, that is the higher energy bands of graphene (fig.\ref{fig:polaritons1}-d). These are inaccessible in graphene due to hybridization with higher bands. Here, the lowest two bands of $s$ symmetry are similar to the $p_z$ bands of graphene. The higher energy orbitals, of $p_x$ and $p_y$ symmetry, form four bands with a remarkable structure consisting of two dispersive bands, which intersect linearly as in graphene, sandwiched   between two additional bands that are flat (Figs.\ref{fig:polaritons1}-e and  \ref{fig:s-p-spectra-edge-states}). The $s$  and $p$ dispersion relations have been measured by photoluminiscence and are displayed in fig.\ref{fig:s-p-spectra}.

\begin{figure}[htbp!]
\begin{center}
\includegraphics[width=5cm]{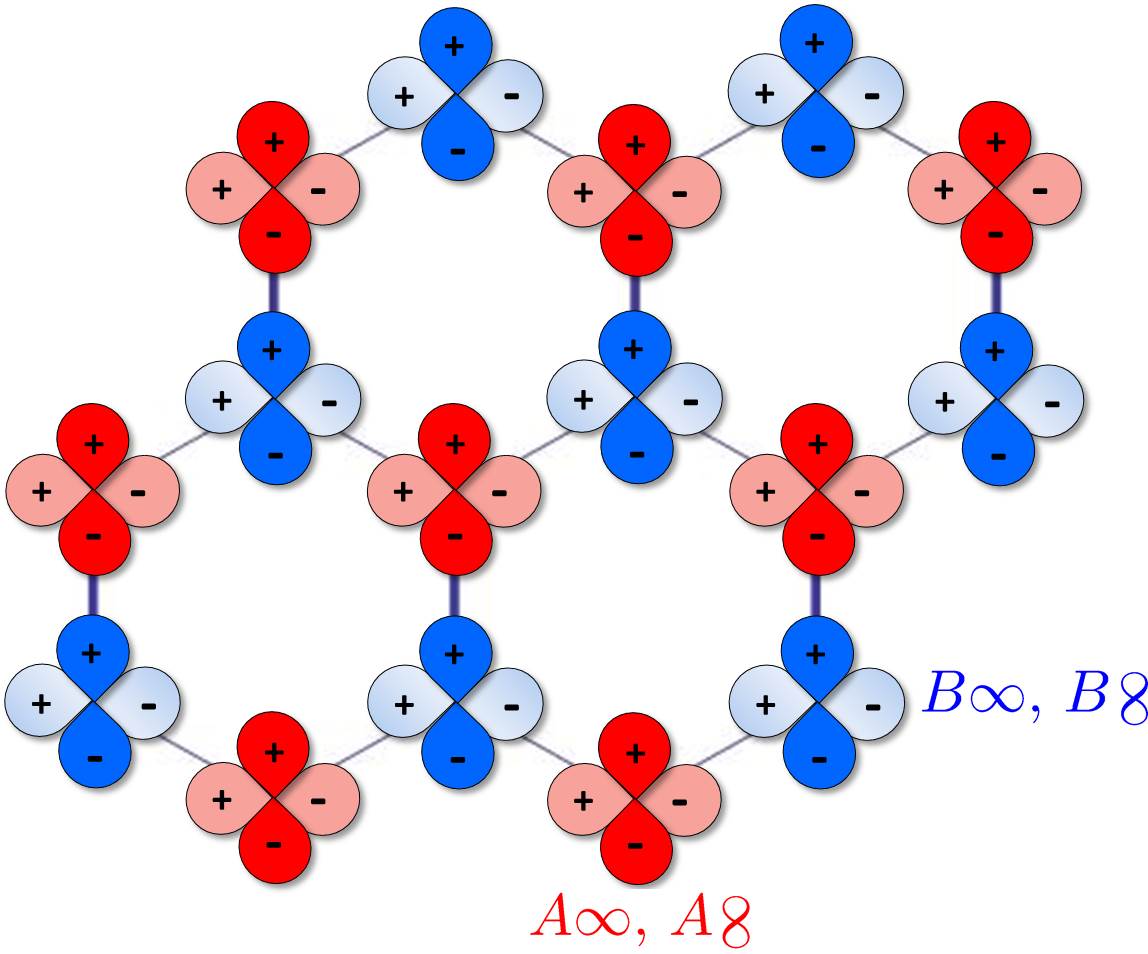}
\end{center}
\caption{\small   $p_x$ and $p_y$ orbitals of $A$ and $B$ sites.}
\label{fig:orbitales-p}
\end{figure}

The $p$-spectrum is well described by a tight-binding model.\cite{Wu}   In the basis of four atomic orbitals ($A$$\infty$, $A$\,\rotatebox[origin=c]{90}{$\infty$}, $B$$\infty$, $B$\,\rotatebox[origin=c]{90}{$\infty$}),
 the $4 \times 4$ Hamiltonian has the form (fig.\ref{fig:orbitales-p})

\be {\cal H}_\k =- t \left(
                    \begin{array}{cc}
                      0 & Q \\
                      Q^\dagger & 0 \\
                    \end{array}
                  \right)
                  \qquad \text{with} \qquad  Q= \left(
                    \begin{array}{cc}
                      f_1 & g \\
                     g & f_2 \\
                    \end{array}
                  \right)
                  \label{H4Q}
                  \ee
with
$f_1= {3 \over 4}(e^{i \k \cdot \a_1} + e^{i \k \cdot \a_2})$, $f_2= \beta + {1 \over 4}(e^{i \k \cdot \a_1} + e^{i \k \cdot \a_2})$ and $g={\sqrt{3} \over 4}(e^{i \k \cdot \a_1} -e^{i \k \cdot \a_2})$.       The numerical factors account for the overlap of the orbitals when they are not facing each other, see fig.\ref{fig:orbitales-p}.
Under anisotropic strain, the anisotropy of hopping parameters is characterized by the parameter $\beta=t'/t$ which enters the expression of $f_2$.
\medskip

\begin{figure}[htbp!]
\begin{center}
\includegraphics[width=8cm]{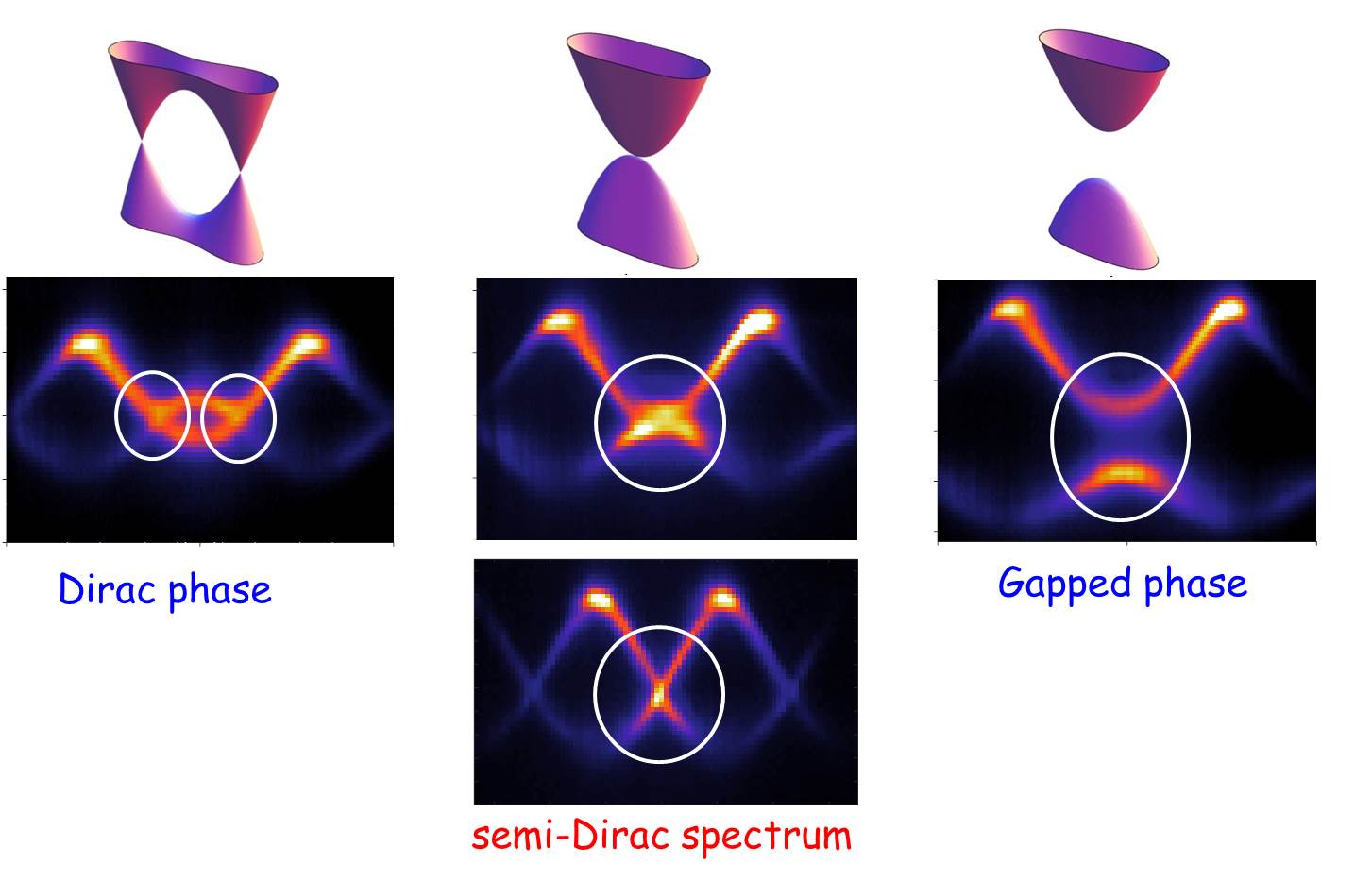}
\end{center}
\caption{\small \it First direct experimental observation of the merging of Dirac points.\cite{c2n3} At the transition, when $\beta=2$, the spectrum in the $s$ band is semi-Dirac: it is linear in one direction and it is quadratic in the other direction.}
\label{fig:semi-dirac}
\end{figure}

\begin{figure}[htbp!]
\begin{center}
\includegraphics[width=8.5cm]{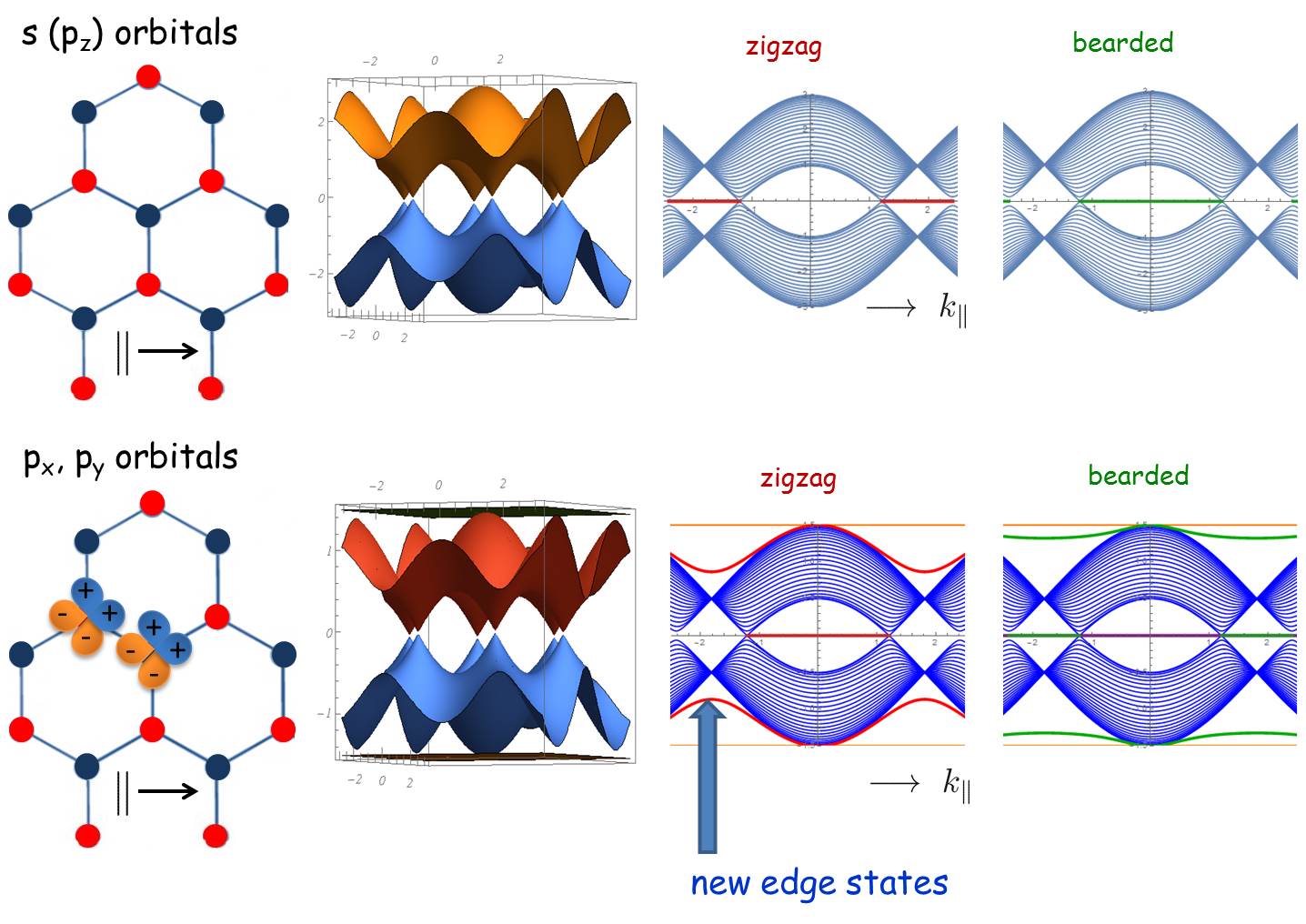}
\end{center}
\caption{Comparison of the $s$ and $p$ spectra. The right panels show the dispersion relation a long the $k_\perp$ direction parallel to the edges. One sees that the "zero energy" (middle of the spectrum) states appear at  complementary edges (zigzag-bearded)   and that new edges states of "finite energy" between the dispersive and the flat bands appear. They have been analyzed in ref.\onlinecite{c2n1}.}
\label{fig:s-p-spectra-edge-states}
\end{figure}

This experimental setup is quite rich.
The fabrication technique\cite{c2n0}
allows to construct honeycomb lattices with arbitrary strain, well described by an anisotropic tight-binding model.  By varying the strain, it has been possible to reach the merging   of Dirac points in the $s$ band, and to measure directly  for the first time the semi-Dirac dispersion relation, as shown in fig.\ref{fig:semi-dirac}.

The physics of the $p$-spectrum exhibits many new features. Recent experiments have probed the physics of edge states and the evolution of the $p$-spectrum under strain.

\begin{figure}[htbp!]
\begin{center}
\includegraphics[width=6cm]{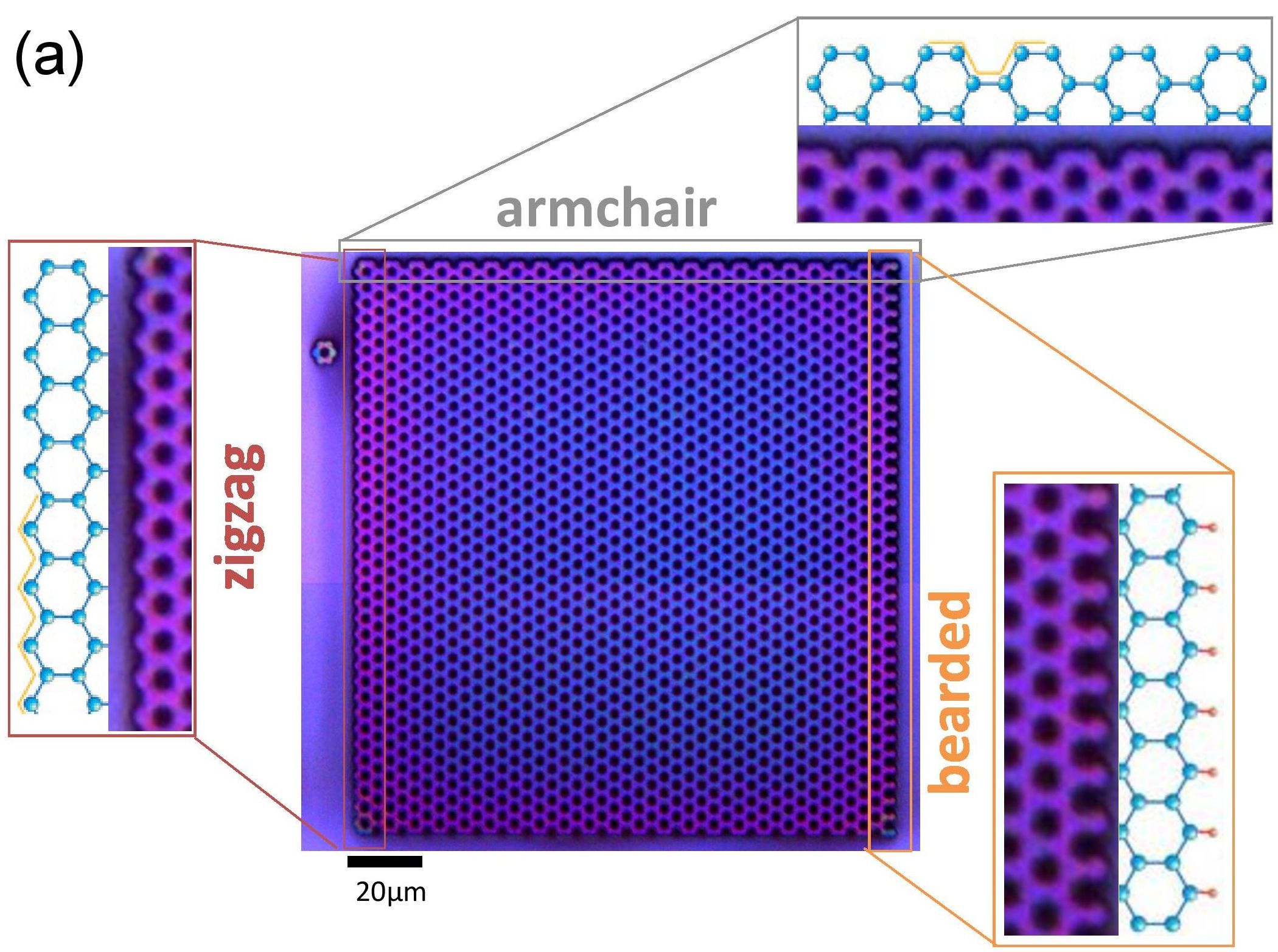}
\end{center}
\begin{center}
\includegraphics[width=9cm]{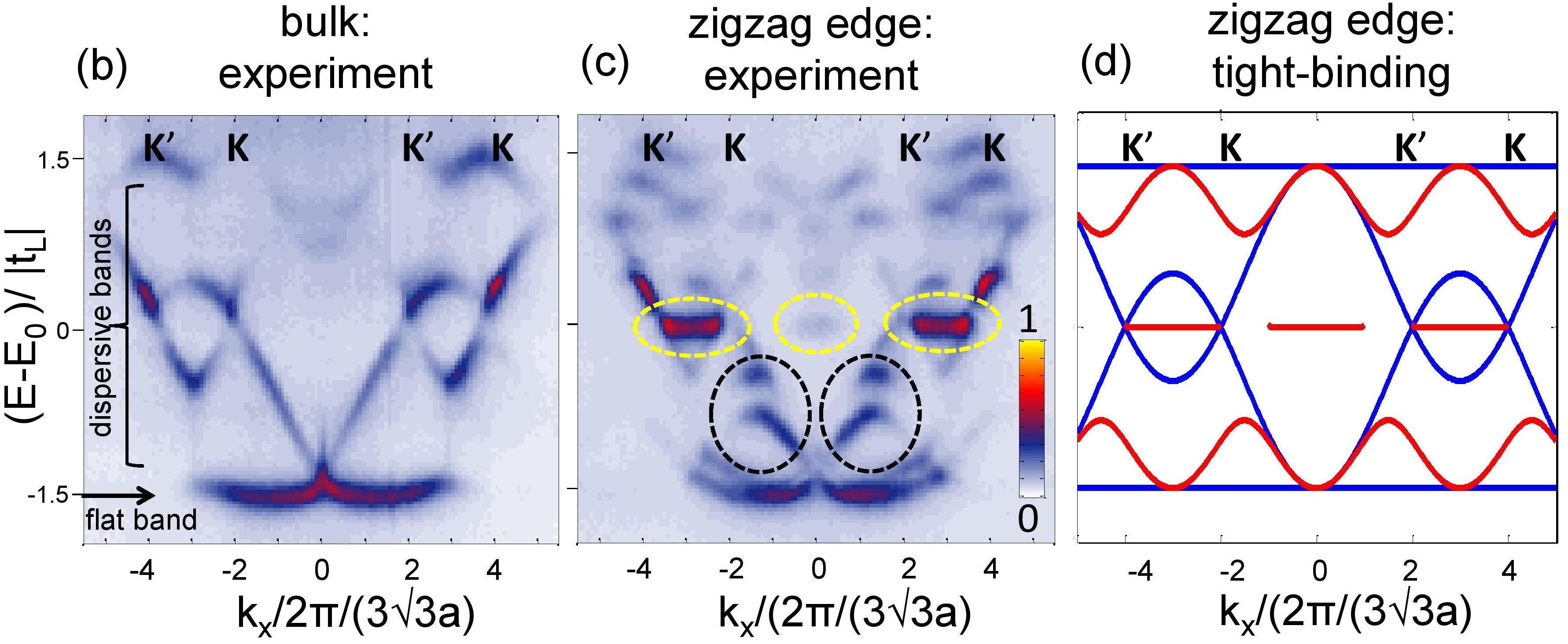}
\end{center}
\caption{\small Top: honeycomb lattice of semiconducting micropillars with three types of edges, zigzag, bearded and armchair. Bottom: photoluminescence spectrum in the bulk and close to a zigzag edge. The tight-binding spectrum exhibits the zero energy edge states and new dispersive states (red) between the bulk dispersive states and the flat bands (blue).}
\label{fig:2_disp}
\end{figure}

Concerning edge states,  it was discovered that   flat "zero energy" (in the middle of the $p$-spectrum) edge states exist at positions in momentum space which are complementary to the ones of the $s$ edge states, see figs.\ref{fig:s-p-spectra-edge-states},\ref{fig:2_disp}. Additionally, new edge states with dispersive character  were found for all types of terminations including armchair. Their  theoretically analysis is presented in ref.\onlinecite{c2n1}.

\begin{figure}[htbp!]
\begin{center}
\includegraphics[width=7cm]{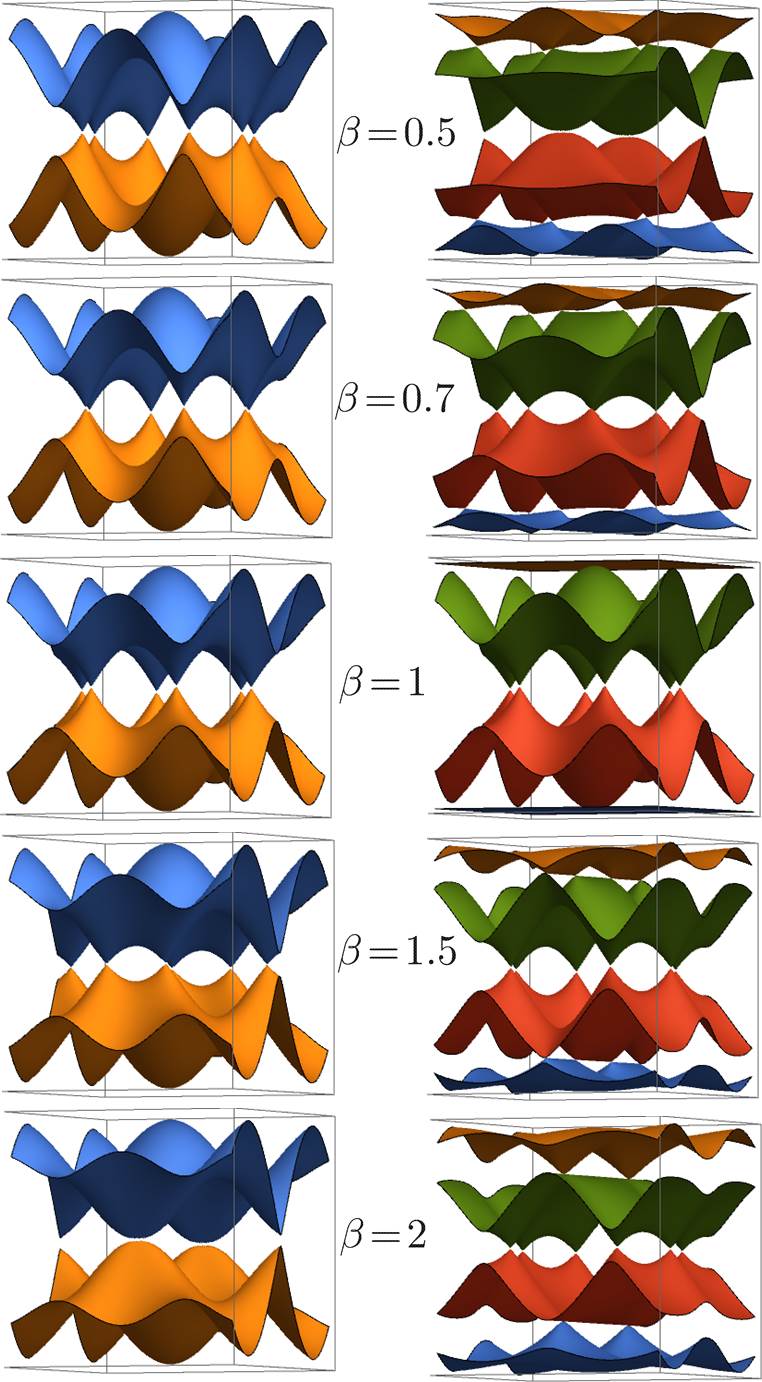}
\end{center}
\caption{\small  Compared evolutions of the $s$ and $p$ spectra under applied strain. When $\beta \neq 1$, two Dirac points emerge from the flat band at the $\Gamma$ point.}
\label{fig:s-p-strain}
\end{figure}

The most recent experiments concern the evolution of the $p$-spectrum under uniaxial strain. Its evolution in the tight-binding model, compared with the one in the $s$-spectrum,  is shown in fig.\ref{fig:s-p-strain}. Several remarkable features have to be discussed. The motion of the Dirac points in the middle of the $p$-spectrum follows an evolution opposite to the one of the $s$-spectrum: merging is reached by {\it decreasing} the strain parameter $\beta=t'/t$ down to the value $\beta=0.5$. The flat bands are deformed as soon as $\beta \neq 1$. A remarkable new feature occurs at the quadratic touching point between the dispersive band and the flat band, either at the top or at the bottom of the spectrum. This quadratic touching point characterized by a winding number $w=2$ splits into a new pair of Dirac points.\cite{c2n2} The scenario of emergence of these points is different from the one described in the above sections where the two Dirac points have {\it opposite} winding numbers and the merging corresponds to the addition rule of the winding numbers $(+,-) \leftrightarrow 0$. In contrast  the two Dirac points here have the {\it same winding number}, and the merging/emergence scenario corresponds to addition rule $(+,+) \leftrightarrow 2$.\cite{Chong,Sun,Dora,Gail2011} This situation is  quite similar to the scenario of splitting of the quadratic touching point observed in bilayer graphene under strain or twist.\cite{McCann,Gail2011}
Moreover these new cones are tilted and  for the first time a pair of critically tilted cones (that is is with a vanishing velocity in one direction) has been observed.\cite{c2n2}
\medskip

To finish this section, we wish to mention a remarkable duality between the $p$-spectrum and the $s$-spectrum. The energy bands --  flat   and dispersive -- of the $p$-spectrum, $\ep^p_{flat}$ and $\ep^p_{disp}$ respectively, are related to those of the $s$-spectrum by the remarkable relation   derived in appendix \ref{appendixB}  (consider here only the positive   energies, in units of the hopping parameters of the $s$ and $p$ bands):

\be  \ep_{flat}^p(\beta) \,  \ep_{disp}^p(\beta) ={3 \over 4} \beta \ep_s\left({1 \over \beta}\right) \label{relationsp} \ee
We have named these $p$-bands, "flat" and "dispersive", but note that the upper (lower) band is flat only when $\beta=1$.
From this duality relation, one can draw several conclusions. When $\beta=1$, the extreme $p$ bands are flat, so that the spectrum of the middle bands is the same as the $s$-spectrum (within renormalization by the hopping parameters). Then one understands immediately the complementarity in the spectra and the edge states described above. A variation of the strain parameter $\beta$ ($\beta$ increase or decrease) has {\it opposite} effects in the $p$ and $s$ spectra.
 Moreover bearded states in the $s$ band are complementary to the zigzag states in the $s$ band and vice-versa.   The wealth of the effects observed in the orbital $p$ band is a motivation for investigation of more sophisticated features in multiband systems.

\section{Phosphorene}
\label{sect.phosphorene}

\begin{figure}[htbp!]
\begin{center}
\includegraphics[width=8cm]{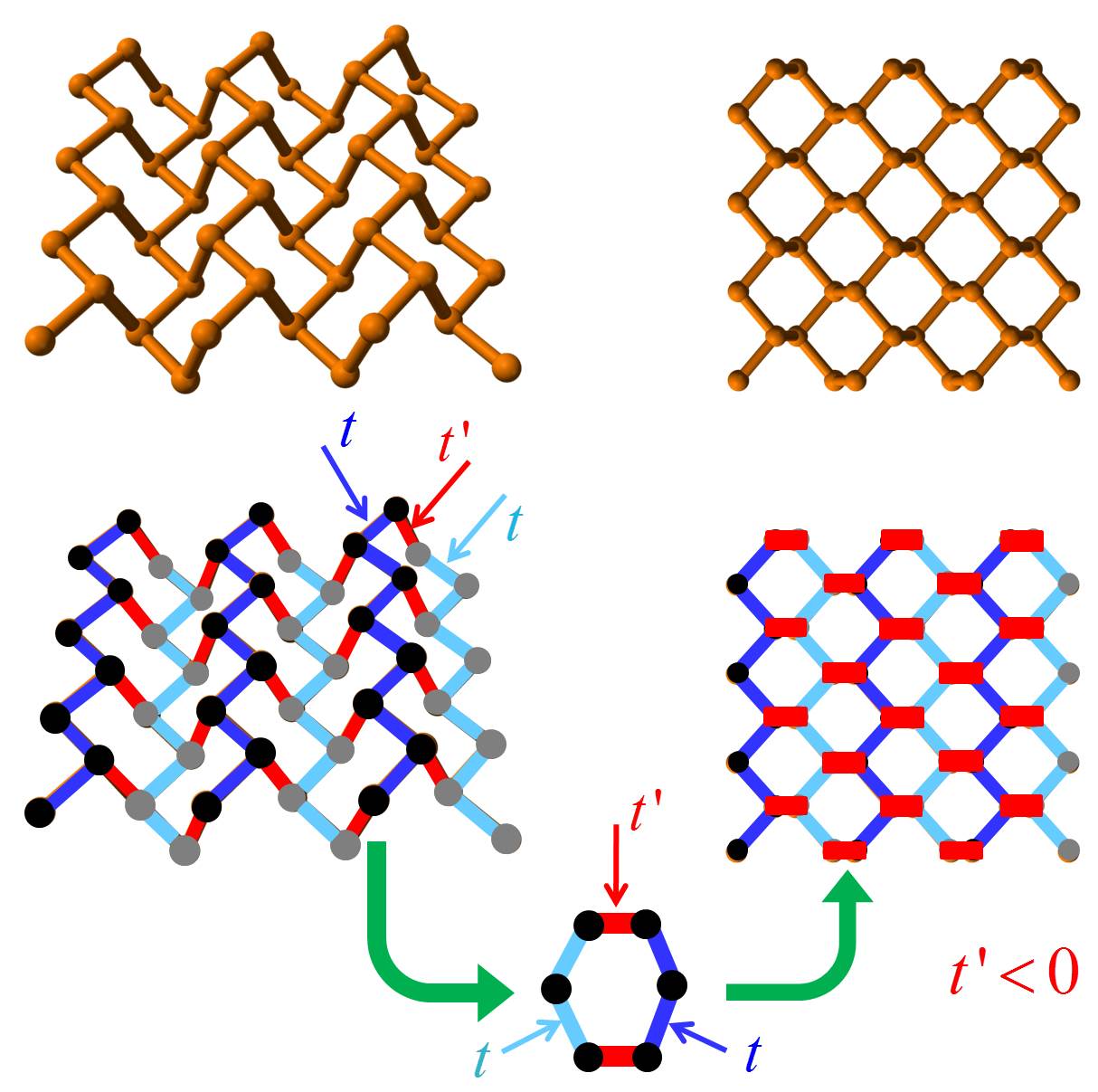}
\end{center}
\caption{\small Like graphene, a  2D crystal of phosphorene may be essentially described by a simple tight-binding model with three hopping integrals $t,t,t'$. In this case $t'$ and $t$ have opposite signs. Left : side view, right : top view.}
\label{fig:phosphorene}
\end{figure}

We  finally discuss the properties of a condensed matter two-dimensional system which bears strong similarities with graphene. Phosphorene is a two-dimensional crystal in which carbon atoms are replaced by phosphorus atoms. Because of $sp_3$ hybridization of $p$ orbitals (instead of $sp_2$ hybridization in graphene), a single sheet of phosphorene is essentially made of two identical planes with horizontal hopping between sites, coupled by an almost vertical hopping (fig.\ref{fig:phosphorene}). At ambient pressure, the electronic spectrum is gapped around the $\Gamma$ point, but under pressure or vertical strain a pair of Dirac points emerges from this point.\cite{Kim,Xiang}
 We show here that this transition follows the scenario of the universal Hamiltonian (\ref{HU}).  Although sophisticated large-scale tight-binding simulations have been developed with nine parameters,\cite{Rudenko} their basic features are captured within our simple toy model with the two parameters of  $t,t'$,  which is  used for strained graphene and that  is also relevant here.

We have shown above that the merging parameter $\Delta_*$ depends on the position of the merging/emergence TRIM point and is given by

\be \Delta_*=t'+ (-1)^p t +(-1)^q t  \ . \ee
In graphene and the realizations discussed above, the two parameters are positive, the transition occurs at the M$_2$ point so that the relevant combination is $\Delta_* = t'-2 t$.  In graphene $t'=t$ so that this parameter is negative. Under strain, $t'$ increases, so that $\Delta_*$ becomes positive and this happens at the topological transition.

\begin{figure}[htbp!]
\begin{center}
\includegraphics[width=8cm]{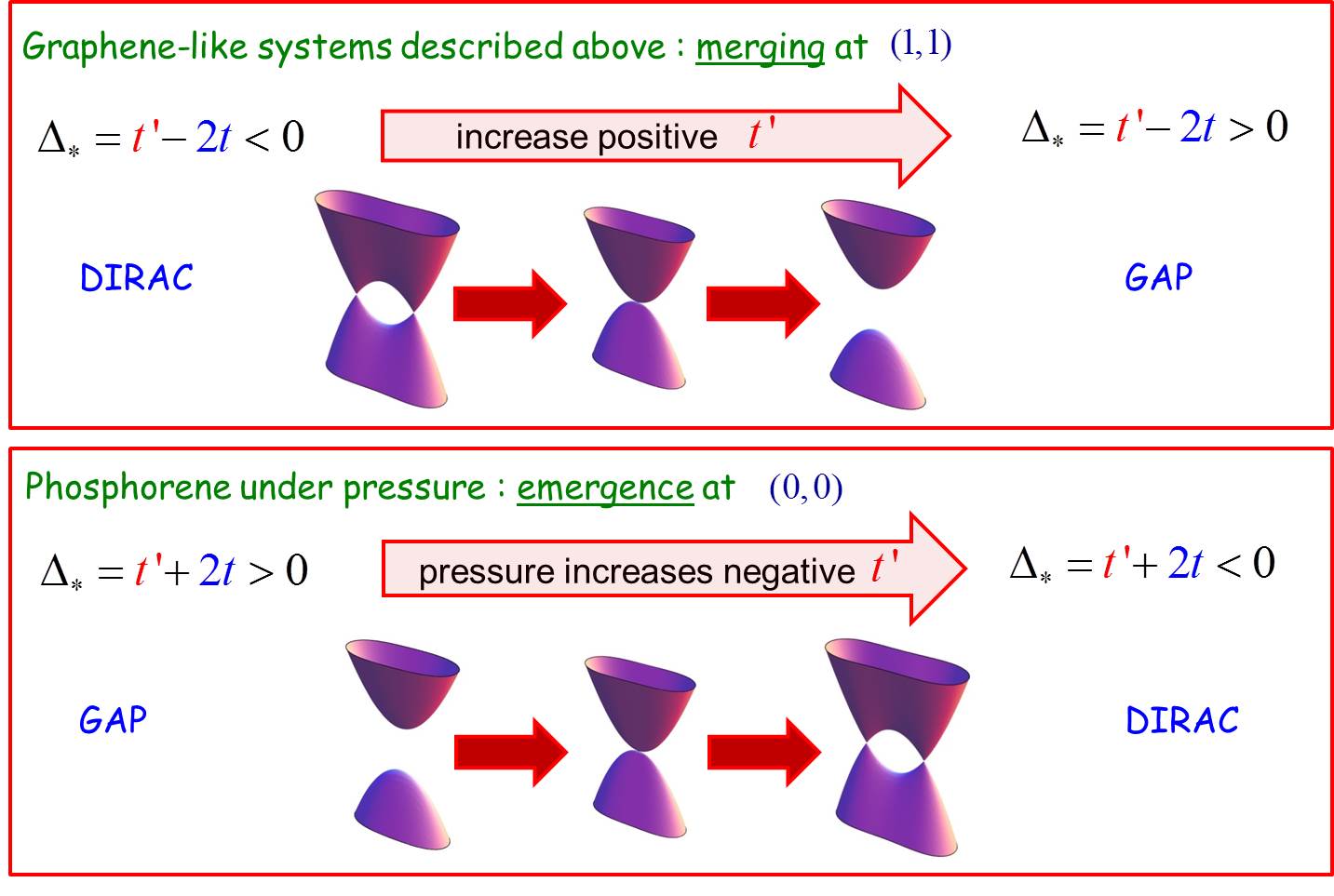}
\end{center}
\caption{\small Similarly to  graphene, a  2D crystal of phospherene may be essentially described by a simple tight-binding model with three hopping integrals $t,t,t'$. In this case $t'$ and $t$ have opposite signs.}
\label{fig:graphene-phosphorene}
\end{figure}

In phosphorene at ambient pressure, the hopping parameters $t$ and $t'$ have opposite signs: $t >0$ and $t' <0$.  The spectrum is gapped which means that $\Delta_* >0$. Under pressure or vertical strain, the  Dirac points emerge  at the $\Gamma$ point ($p=q=0$) and this phenomenon is therefore driven by the parameter $\Delta_* =t'+2 t$ which becomes negative. The reason is that the amplitude of negative $t'$ increases under pressure.
This emergence of Dirac points is very well described by the universal Hamiltonian. The calculated evolution of the Landau level spectrum in a complete tight-binding description of phosphorene is quite similar to the expected spectrum obtained from the universal Hamiltonian (compare fig.\ref{fig:landau-spectrum} and fig.7 of ref.\onlinecite{Pereira2}).

The two scenarios of merging or emergence in graphene-like structure and phosphorene are summarized in fig.\ref{fig:graphene-phosphorene}. We conclude that the merging/emergence of Dirac points in strain (artificial graphene) or phosphorene follow the same universal scenario.

\section{Conclusion}

We have described several physical systems where the excitation spectrum exhibits a pair of Dirac cones, such as graphene. The flexibility of these different systems allows one to vary hopping parameters so that these Dirac points can be manipulated: they can move, merge or emerge at special points of high symmetry of the reciprocal lattice.

Each Dirac point is characterized by a winding number $w= \pm 1$, and a pair of  Dirac points can merge following two alternative scenarios : either they have opposite "charge" and this corresponds to the scenario discussed in this paper, $(+-) \leftrightarrow 0$, or they have identical charge and this corresponds to a second merging scenario, $(++) \leftrightarrow 2$, described by  another effective Hamiltonian, and which is not discussed in this paper. For a discussion on this second scenario, see the  references \onlinecite{Gail2012,Poincare}.
   The merging between two Dirac cones is thus a topological transition that may be described by two distinct universality classes, according to whether the two cones have opposite or identical topological charges. We have recently found a system in which the two classes may be observed:
indeed, photoluminescence experiments on the $p$-spectrum of a strained honeycomb polariton lattice have shown a very rich variety of Dirac points. Both   types of merging have been observed.\cite{c2n2}
In particular we have highlighted the existence of a pair of Dirac points emerging from the touching point between a quadratic band and a flat band, following the scenario $(+,+) \leftrightarrow 2$.   It turns out that under further strain, this pair of Dirac points may merge at another TRIM following the scenario $(+,-) \leftrightarrow 0$.\cite{c2n2,Montambaux:18}
This apparent contradiction is solved by the fact
that the winding number is actually defined around a unit vector on the Bloch sphere and that
this vector rotates during the motion of the Dirac points.\cite{Montambaux:18} These new results open the way for future investigations on  the production of Dirac points, their geometric properties  and their fate under appropriate variations of external parameters, especially in  multiband systems.

\acknowledgments

This work has benefited from collaborations and discussions with A. Amo, M. Bellec, C. Bena, J. Bloch, P. Delplace, R. de Gail, P. Dietl, J.-N. Fuchs, M.-O. Goergig, U. Kuhl,  L.-K. Lim, M. Mili\'cevi\'c, F. Mortessagne, T. Ozawa, F. Pi\'echon, D. Ullmo. Special thanks to J.-N. Fuchs and M.-O. Goergig for a careful rereading of the manuscript.

\appendix

\section{A remark on two representations of a Bloch function}
\label{appendixA}

In the tight-binding picture, the eigenstates, solutions of the Hamiltonian (\ref{HTB}) are a combination of atomic orbitals satisfying Bloch theorem. They can be written in the alternative forms\cite{Bena}

\be |\psi_{\k} \rangle = {1 \over \sqrt{N}} \sum_j e^{i \k \cdot \R_j} \left( c^A_\k |\varphi^A_j \rangle + c^B_\k |\varphi^B_j \rangle \right)
 \ee
or
\be |\psi_{\k} \rangle = {1 \over \sqrt{N}} \sum_j  \left( e^{i \k \cdot \R_j^A}  d^A_\k |\varphi_j^A \rangle + e^{i \k \cdot \R_j^R} d^B_\k  |\varphi_j^B \rangle \right)
 \ee
 where $\varphi^A_j$ and $\varphi^B_j$ are atomic orbitals on sites $(j,A)$ and $(j,B)$.
This is the same eigenstate, but these two representations are slightly different. In the first expression, the phase factor in attached to the elementary cell and is common to the $A$ and $B$ sites. This is for example the form found in the Ashcroft and Mermin book (eq. 10.26).\cite{Ashcroft} In the second writing, each phase factor is attached to the position of each atom, $A$ or $B$. This is for example the notation found in the Wallace paper.\cite{Wallace} By choosing $\R_j=\R_j^A$, we have obviously $c_\k^A=d_\k^A$ and $c_\k^B=d_\k^B e^{i \k  \cdot (\R^B-\R^A)}= d_\k^B e^{i \k \cdot {\bs \delta}_3}$, $\delta_3$ being the $AB$ distance (fig.\ref{fig:vecteurs}).   The two representations of the same eigenfunction are of the  form

\begin{eqnarray}  \psi(\r)&=& e^{i \k \cdot \R} \, u_{I,\k}(\r)  \\
 \psi(\r)&=& e^{i \k \cdot \r}\,  u_{II,\k}(\r)  \end{eqnarray}
where $\r$ is the position and $\R$ is a vector of the Bravais latttice, position of the elementary cell in the lattice. Note that the function  $u_{I,\k}(\r)$ has the periodicity of the reciprocal lattice, while $u_{II,\k}(\r)$, which is the most common representation, does not.
 \medskip

 In the tight-binding model (\ref{HTB}), the wave function is defined only on the $A$ and $B$ sites so that the normalized wave functions can be written in the spinorial form:
\begin{eqnarray} \text{either} \qquad u_{I,\k}&=& \left(
            \begin{array}{c}
              c^A_\k \\
              c^B_\k \\
            \end{array}
          \right)=
{1 \over \sqrt{2}} \left(
                               \begin{array}{c}
                                 1  \\
                                \pm e^{i \phi_{I,\k}} \\
                               \end{array}
                             \right)
\\
\text{or} \qquad
  u_{II,\k}&=& \left(
            \begin{array}{c}
              d^A_\k  \\
              d^B_\k \\
            \end{array}
          \right)=
{1 \over \sqrt{2}} \left(
                               \begin{array}{c}
                                 1  \\
                               \pm  e^{i \phi_{II,\k}} \\
                               \end{array}
                             \right) \ .
\end{eqnarray}
They are solutions of the $2 \times 2$ Hamiltonian
\be {\cal H}_{i,\k}= \left(
                   \begin{array}{cc}
                     0 & f_{i,\k} \\
                     f^*_{i,\k} & 0 \\
                   \end{array}
                 \right)
                 \ee
with
\begin{eqnarray}\text{either}  \qquad f_{I,\k} &=& - t( 1 + e^{i \k \cdot \a_1} + e^{i \k \cdot \a_2} )  \\
\text{or} \qquad   f_{II,\k}&=& - t(   e^{i \k \cdot \d_3} + e^{i \k \cdot \d_1}  + e^{i \k . \d_2} )
\end{eqnarray}
and the relative phase between the two components of the wave function is given by
$\phi_{i,\k}= \arg( f_{i,\k}^*)$, so that

\be \phi_{II,\k}=\phi_{I,\k} -   \k  \! \cdot \! \d_3  \ . \ee

\begin{figure}[htbp!]
\begin{center}
\includegraphics[width=4.2cm]{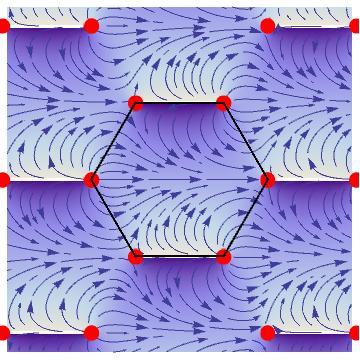}\hfill\includegraphics[width=4.2cm]{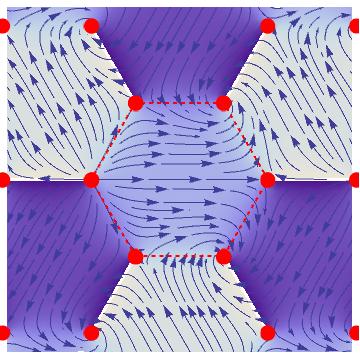}
\end{center}
\caption{\small  Representation of the phases $\phi_{I,\k}$ (left) and $\phi_{II,\k}$ (right), via the vector field $(\cos \phi,\sin \phi)$. The phase $\phi_{I,\k}$ has the periodicity of the reciprocal lattice. The phase $\phi_{II,\k}$ has a triple periodicity. This phase has been measured directly in interferometry experiments with a honeycomb lattice of cold atoms\cite{Schneider,Tarnowski,Lim:15} and in ARPES experiments on graphene (fig.\ref{fig:arpes}).\cite{Mucha}}
\label{fig:phases}
\end{figure}

In the first representation, the phase $\phi_{I,\k}$ has the periodicity of the reciprocal lattice :  $\phi_{I,\k + \G} = \phi_{I,\k } $, see fig.\ref{fig:phases}. Obviously, because of the linear term $\k \! \cdot \!  \d_3$ related to the $A-B$ distance, the phase $\phi_{II,\k }$ does not have this periodicity but a triple periodicity (fig.\ref{fig:phases}).

\begin{figure}[t!]
\begin{center}
\includegraphics[width=6cm]{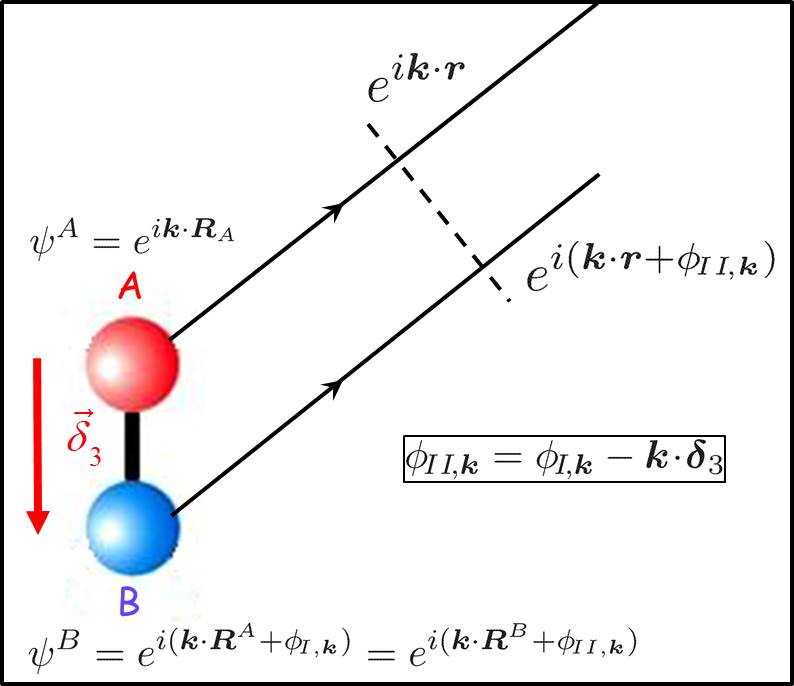}
\end{center}
\caption{\small The waves emitted by the sites of the honeycomb lattice can be described by a "two-source" interference pattern.\cite{Mucha}}
\label{fig:emitters}
\end{figure}

\begin{figure}[htbp!]
\begin{center}
\includegraphics[width=7cm]{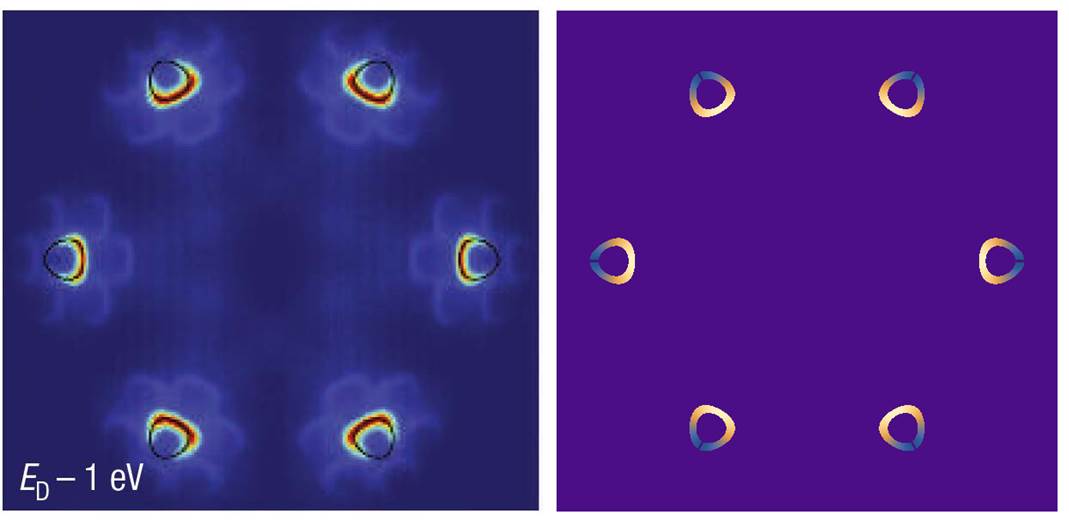}
\end{center}
\caption{\small Left : ARPES measured intensity\cite{Mucha}. Right : Intensity given by   $\cos^2 {\phi_{II,\k}  \over 2}$. }
\label{fig:arpes}
\end{figure}

What is the physical interpretation of these two phases $\phi_I$ and $\phi_{II}$ ?  The periodic honeycomb lattice can be represented by a two-source emitter\cite{Mucha} corresponding to the two sublattices, see fig.\ref{fig:emitters}. $\phi_{I,\k}$ is the relative phase between the two emitters, that is the phase delay between the waves emitted by the two sites. It does not account for the relative position between these two sites. If $\k \! \cdot \! \r$ represents the phase emitted by atoms $A$ at any point $\r$, the combination $\k \!\cdot\! \r +\phi_{II,\k}$ represents the phase emitted by atoms $B$ at the same point (fig.\ref{fig:emitters}). The signals emitted by the two sources construct an interference pattern which depends on the relative position between the two sources. At a position $\r$ outside the sample, the total electronic wave has the form $e^{ i \k   \cdot   (\r -\R_A)}  \psi_A + e^{  i \k   \cdot  (\r - \R_B)}  \psi_B = (1 + e^{i \phi_{II,\k} })
  e^{i \k  \! \cdot \! \r}$ so that the intensity $I(\k)$ is proportional to $\cos^2 {\phi_{II,\k}\over 2} $, as measured in ARPES experiments, see fig.\ref{fig:arpes}.

 In this paper, we have used the representation I   for  the writing of the Hamiltonian (\ref{Heff}) in its general form (\ref{fgeneral})  which has the periodicity of the reciprocal lattice. It is also used for describing geometric properties of the Hamiltonian and the wave functions, the relation between existence of edge states and the Zak phase.\cite{Delplace:11} The phase $\phi_{II,\k}$ is directly measured in interferometry experiments on cold atoms\cite{Schneider,Lim:15,Tarnowski} or in ARPES experiments,\cite{Mucha} because the interference pattern depends on the relative position $\d_3$ between atoms $A$ and $B$. This phase does not have the periodicity of the reciprocal lattice.

\section{Proof of eq.(\ref{relationsp})}
\label{appendixB}

 This equation relates the dispersion relation of the $p$ bands to the dispersion relation of the $s$ band. The $p$-spectrum consists in four bands of energies named $\pm \ep_{flat}$ and $\pm \ep_{disp}$. Note that the two "flat" bands are strictly flat only when $\beta=1$.

Consider first the $s$-spectrum. It is given by $\pm \ep_s(\beta)$ with
$\ep_s(\beta)= |f_\k|$,   where $f_\k$ is given by eq.(\ref{fdektoy}). Therefore in units of $t$, it can be written as
\be \ep_s(\beta)= | \beta + z_1 + z_2 |\ , \ee
where $z_i= e^{i \k \cdot \a_i}$.

Consider now the eigenvalues of the $p$-Hamiltonian (\ref{H4Q}). Simple algebra shows that the eigenvalues are solutions of the equation
\be \ep^4 - S \ep^2 + P = 0 \ , \ee
 with
 \be \quad  S= \text{tr}(QQ^\dagger) \quad ,  \quad P= \text{det}(QQ^\dagger)= |\text{det}\, Q|^2  \ ,\ee
so that
\be   \ep_{flat}^p \,  \ep_{disp}^p = |\text{det}\, Q|  \ . \ee
Writing the elements of the matrix $Q$ as $f_1={3\over 4}(z_1+z_2)$, $f_2= \beta + {1 \over 4} (z_1+z_2)$ and $g={\sqrt{3} \over 4} (z_1-z_2)$, we find
\begin{eqnarray}
   \ep_{flat}^p(\beta) \,  \ep_{disp}^p(\beta) &=& {3 \over 4} \beta \, \left|{z_1 z_2 \over \beta}+ z_1 + z_2 \right|\nonumber \\
    &=&{3 \over 4} \beta \, \left|{1\over \beta} + z_1 + z_2  \right| \end{eqnarray}
since $|z_i|=1$. This proves eq.(\ref{relationsp}).

\end{document}